# Identifying Crisis-Critical Intellectual Property Challenges during the Covid-19 Pandemic: A scenario analysis and conceptual extrapolation of innovation ecosystem dynamics using a visual mapping approach

Alexander Moerchel[1, 2, *], Frank Tietze[1], Leonidas Aristodemou[1], Pratheeba Vimalnath[1]

[1.] Innovation and IP Management (IIPM) Lab, Centre for Technology Management (CTM), Institute for Manufacturing (IfM), Department of Engineering, University of Cambridge, Cambridge, United Kingdom

[2] Lufthansa Technik, Hamburg, Germany

* Please contact the corresponding author: amm21@cam.ac.uk

UNIVERSITY OF CAMBRIDGE
Department of Engineering

IfM Centre for Technology Management

# Identifying Crisis-Critical Intellectual Property (IP) Challenges during the Covid-19 Pandemic: A scenario analysis and conceptual extrapolation of innovation ecosystem dynamics using a visual mapping approach

Alexander Moerchel[1,2,*], Frank Tietze[1], Leonidas Aristodemou[1], Pratheeba Vimalnath[1]

## Abstract

The Covid-19 pandemic exposed firms, organisations and their respective supply chains which are directly involved in the manufacturing of products that are critical to alleviating the effects of the health crisis, collectively referred to as the Crisis-Critical Sector, to unprecedented challenges. Firms from other sectors, such as automotive, luxury and home appliances, have rushed into the Crisis-Critical Sector in order to support the effort to upscale incumbent manufacturing capacities, thereby introducing Intellectual Property (IP) related dynamics and challenges. We apply an innovation ecosystem perspective on the Crisis-Critical Sector and adopt a novel visual mapping approach to identify IP associated challenges and IP specific dynamic developments during and potentially beyond the crisis. In this paper, we add methodologically by devising and testing a visual approach to capturing IP related dynamics in evolving innovation ecosystems and contribute to literature on IP management in the open innovation context by proposing paraground IP as a novel IP type. Finally, we also deduce managerial implications for IP management practitioners at both incumbent firms and new entrants for navigating innovation ecosystems subject to crisis-induced dynamic shifts.

*Affiliations:*
[1] Innovation and IP Management (IIPM) Lab, Centre for Technology Management (CTM), Institute for Manufacturing (IfM), Department of Engineering, University of Cambridge
[2] Lufthansa Technik, Hamburg, Germany
* Corresponding author

*Emails:*
Alexander Moerchel (amm21@cam.ac.uk), Frank Tietze (frank.tietze@cam.ac.uk), Leonidas Aristodemou (la324@cam.ac.uk), Pratheeba Vimalnath (pv302@cam.ac.uk)

## Table of Contents

## 1.0 Introduction

The Covid-19 pandemic has dramatically affected the global economy since its outbreak in the Hubei province in China in December 2019 (Sohrabi *et al.*, 2020) and during the subsequent spread throughout the Western Pacific, Europe, the Americas and the Eastern Mediterranean (World Health Organisation, 2020b). Particularly those firms and organisations appeared to be under enormous pressure, which are directly involved in the research, development and the production of crisis-critical products that are relevant to managing the crisis of the Covid-19 pandemic through prevention, diagnosis and treatment, such as Personal Protection Equipment (PPE), diagnostic test kits, drugs potentially relevant for treating Covid-19 patients, ventilators, vaccines, and so on.[1]

Not only are incumbent firms in the CC-Sector under particular pressure to provision the required capacities for the manufacturing of CC-Products, but also face enormous challenges to deal with changes to their ecosystem as suddenly new entrants rush in to support capacity building. To succeed, new entrants, incumbents or all of them together have been witnessed to be innovative during this pandemic (Tietze *et al.*, 2020). While Intellectual Property (IP) often does not feature prominently in discussions related to the current Covid-19 induced crisis, various papers indicate that IP has a role to play for ending the pandemic (Azoulay and Jones, 2020; Contreras *et al.*, 2020; Tietze *et al.*, 2020). Adopting a visual mapping approach in this paper, we attempt to clarify some of the industrial dynamics induced by the pandemic and uncover IP associated challenges that affect CC-Sector actors, particularly incumbents CC-Product manufactures and new entrants, which then helps us to understand how different actors can use IP to react to current dynamics and position themselves for a time after the pandemic.

[1] The incumbent firms and organisation are, together with their respective supply chains, henceforth collectively denoted as the Crisis-Critical Sector or CC-Sector (Tietze *et al.*, 2020), while products are referred to in the remainder of this paper as Crisis-Critical Products or CC-Products (Elsen *et al.*, 2020).

When the pandemic struck, on one side, incumbent CC-Sector firms and organisations faced a positive demand shock for CC-Products by the exponentially growing number of patients (Phua *et al.*, 2020), doctors and hospitals treating them, public health systems, and non-governmental organisations (NGOs), exceeding their inherent manufacturing capacities. On the other hand, these firms have to cope with global supply networks that are disrupted by heterogenous government responses and national shutdowns caused by measures reducing contact rates in general populations (Ferguson *et al.*, 2020; Nicola *et al.*, 2020). For instance, India, which is the biggest producer of generics effective in the treatment of milder symptoms of Covid-19, such as Paracetamol, experienced a disruption in the supply of necessary active pharmaceutical ingredients from China during its industrial shutdown induced by the Covid-19 pandemic (Oxford Business Group, 2020).

In a recent analysis of research, development and manufacturing upscaling activities during the initial phases of the Covid-19 pandemic, Tietze et al. (2020) summarise current dynamics in the CC-Sector, which they argue would lead to changes of industrial organisation and highlight the importance of IP related challenges. More specifically, they conclude that new entrants being called into the CC-Sector to support incumbent firms alleviating manufacturing capacity bottlenecks for CC-Products raises questions with regards to the handling of IP underlying CC-Products, henceforth denoted as Crisis-Critical IP or CC-IP, that need to be addressed sooner rather than later during an unfolding global health crisis in order to avoid them becoming hold-ups in the effort to meet the positive demand shock for CC-Products.

In this study, we build on Tietze et al. (2020) and answer to their call for a more detailed understanding of IP-related challenges in the CC-Sector during the Covid-19 pandemic. In this paper, we visually represent changes in the CC-Sector that take place during the pandemic and discuss and visualise those that can be expected after the pandemic. For this exercise, we conceptualise the CC-Sector as an innovation ecosystem with incumbents at its centre and capture dynamics that emerge when new entrants join incumbents from outside the CC-Sector to build manufacturing capacity for CC-Products. We further identify potential IP associated challenges from the incumbents' perspective for each phase of the pandemic. In so doing, we explore how loci of innovation, as well as IP ownership and usage, can be graphically represented in order to visually capture the inherent dynamics in an innovation ecosystem and changing IP arrangements specifically.

For visually representing the scenarios described by Tietze et al. (2020), we apply and extend a visual language for depicting ecosystem dynamics proposed by Moerchel et al. (2021) by introducing visual representations to capture IP arrangements. Furthermore, we model Tietze et al.'s (2020) scenario in which industrial manufacturers enter CC-Sector from outside, therein considered as Type 1 New Entrants. These are either called in or voluntarily enter the CC-Sector to help incumbent firms with increasing manufacturing capacities for CC-Products in response to the Covid-19 pandemic crisis situation. These Type 1 New Entrants can be classified in four groups: (1) manufacturers with spare valuable resources and capabilities, (2) tech giants with diverse capabilities and rich resources, (3) firms with relevant expertise, resources and competence, or (4) agile firms with a particular set of skills. Tietze et al. (2020) furthermore identify three distinct strategies, which Type 1 New Entrants follow to enter the CC-Sector, that is to access the necessary CC-IP typically owned by incumbent firms in the CC-Sector for the production of CC-Products. These entry strategies include (A) insouciantly adapting features of incumbent CC-Products by copying or reverse-engineering without conducting freedom-to-operate analysis, (B) designing CC-Products from the ground up using own competence and expert advice, thereby potentially developing their own CC-IP, but without ensuring proper freedom-to-operate also risking infringement claims, or (C) teaming up with incumbent CC-Product manufacturer thereby licensing in CC-IP. Table 1 below summarises and aligns the Type 1 New Entrant groups to entry strategies into the CC-Sector and provides empirical examples for each.

Table 1: Alignment of Type 1 New Entrant group to entry strategy (synthesised from Ambrosio, 2020; Tietze et al., 2020)

<table>
<tr><th>Strategy</th><th>Type 1 New Entrant Group</th><th>Example New Entrants</th><th>Example CC-Product</th><th>Reference</th></tr>
<tr><td rowspan="5">(A) Insouciantly Adopting</td><td rowspan="5">(4) Agile manufacturers with particular and complementary skills to the CC-Sector</td><td>Formula 1 Teams</td><td>Ventilator</td><td>(Richards, 2020)</td></tr>
<tr><td>LVHM, Estee Lauder and L'Oréal</td><td rowspan="2">Hand Sanitizer</td><td>(Pays, 2020)</td></tr>
<tr><td>BrewDog and Old Forth Distillery</td><td></td></tr>
<tr><td>Parkdale Mills Inc</td><td rowspan="2">PPE (e.g. face shields, masks, hospital gowns)</td><td>(Ma, 2020)</td></tr>
<tr><td>Prusa3D, Formlabs and Stratasys</td><td>(Manero et al., 2020)</td></tr>
<tr><td rowspan="2">(B) Designing from Scratch</td><td rowspan="2">(3) Firms with relevant expertise, resources and competence</td><td>Dyson in co-operation with The Technology Partnership</td><td>Dyson CoVent Ventilator</td><td>(Jack, 2020)</td></tr>
<tr><td>Roche</td><td>Roche's cobas Sars-CoV-2 test</td><td>(F. Hoffmann-La Roche AG, 2020)</td></tr>
<tr><td>(C) Teaming Up</td><td>(1) Manufacturers with spare valuable resources and capabilities<br>(2) Tech giants with diverse capabilities and rich resources</td><td>GM in collaboration with Ventec Life Systems</td><td>Ventilator</td><td>(Brooks and Flores, 2020)</td></tr>
</table>

With this paper we first contribute methodologically by extending and adapting a newly developed visual approach, namely the Standardised Visual Ecosystem Language, to innovation ecosystem analysis in the crisis use case, thereby expanding the toolbox of IP strategy practitioners navigating dynamically evolving innovation ecosystems. Subsequently, we apply the adapted and extended standardised visual approach, which results in proposing paraground IP as a new type of IP that plays a role in the open innovation paradigm. Finally, this paper contributes a set of managerial implications and recommendations for IP strategists, practitioners and decision-makers at both incumbent and new entrant firms experiencing crisis-induced dynamic shifts in their innovation ecosystem.

The remainder of this paper introduces the research approach including a description of the standardised visual language for mapping innovation ecosystems (Chapter 2), summarises and discusses the results from mapping the CC-Sector innovation ecosystem for each of the three new entrant strategies during the pre-pandemic, pandemic and post-pandemic phases (Chapter 3) before closing with a conclusion, listing limitations of this study and recommending directions for future work (Chapter 4).

## 2.0 Research Approach

In order to delineate IP-related challenges, risks and uncertainties in the CC-Sector innovation ecosystem during and after the Covid-19 induced crisis, we adopt a qualitative exploratory research strategy. We perform an empirically informed, detailed analysis of past and current events coupled with a conceptually guided extrapolation into the future of each of the three entry strategies empirically identified by Tietze et al. (2020). The three entry strategies (see Table 1) for large manufacturing firms called in or voluntarily joining the CC-Sector to help the effort of upscaling the production of CC-Products during the crisis are: (A) Insouciantly Adopting, (B) Designing from Scratch and (C) Teaming up. Specifically, we apply the Nominal Group Technique (NGT) to: (i) adapt an existing ecosystem mapping procedure and visual language, referred to as the Standardised Visual Language (SVEL), to empirically analyse past and current events related to the Covid-19 pandemic; and (ii) synthesise future developments in the CC-Sector innovation ecosystem for each entry strategy scenario guided by received concepts in competitive dynamics. NGT[2] represents a structured, direct interaction procedure involving a group of subject matter experts (Van De Ven and Delbecq, 1974; Robson and McCartan, 2017). The composition of the supporting expert panel is summarised in Table 2, along with the areas of expertise and affiliation of each NGT member.

### 2.1 Nominal Group Technique (NGT) research methodology

The NGT procedure is implemented iteratively in several steps[3] (Van De Ven and Delbecq, 1974). Figure 1 summarises the NGT-based research method described below. Firstly, using the previously developed SVEL, the NGT expert panel coordinator prepares maps of the CC-

[2] The Nominal Group Technique (NGT) is chosen because of its proven capability to achieve more accurate judgement about future outcomes in the face of uncertainty, compared to statistically aggregated independent judgement, such as questionnaires (Woudenberg, 1991).

[3] We adjust the Nominal Group Technique (NGT) to account for the Covid-19 induced NPIs, such as travel restrictions and widespread closures of research facilities. This requires the NGT experts to meet remotely via digital means.

Sector innovation ecosystem, for each of the three entry strategies described in Table 1, during the pre-pandemic, pandemic and post-pandemic phases of the Covid-19 induced crisis.

Table 2: Composition of NGT expert panel, areas of expertise and affiliation

| NGT Member | Affiliation | Areas of Expertise | Sector |
|---|---|---|---|
| NGT Member 1 (Leader/Coordinator) | Academic and Industry | • Negotiation of Purchasing and Licensing Contracts<br>• Innovation and IP Management | • Manufacturing and Management<br>• Aerospace MRO |
| NGT Member 2 | Academic | • Innovation and IP Management<br>• Open IP Strategies & IP Pledges<br>• Technology Licensing | • Biotechnology and Bioeconomy<br>• Economics<br>• Manufacturing and Management |
| NGT Member 3 | Academic | • Economics of IP<br>• Corporate Strategy<br>• Technology and Innovation Management | • Economics<br>• Manufacturing and Management |
| NGT Member 4 | Academic | • Open IP Approaches<br>• Economics of Patents and Patent Transactions<br>• Technology Management | • Biotechnology and Bioeconomy<br>• Manufacturing and Management |
| NGT Member 5 | Academic | • IP Analytics and IP Intelligence<br>• Strategic Decision-making<br>• Artificial Intelligence | • Manufacturing and Management |
| NGT Member 6 | Academic | • Open Approaches to IP<br>• Biotechnology and bioeconomic policy | • Biotechnology and Bioeconomy |

Subsequently, the digitally transcribed innovation ecosystem maps are circulated to the NGT expert panel for individual review to accurately capture IP dynamics in the CC-Sector innovation ecosystem as the Covid-19 crisis progresses from the pre-pandemic through the pandemic to the post-pandemic phase. During the collective group review follow-up, the coordinator goes through each entry strategy scenario and every NGT member has the opportunity to request clarifications, present their feedback and discuss their view. The coordinator carefully records the feedback and views of each NGT member and summarises them for subsequent circulation to the expert panel with a final request for confirmation. On the basis of this consolidated feedback, the coordinator revises the innovation ecosystem maps and resubmits them to the NGT expert panel. This procedure is repeated until the NGT expert panel is unanimously satisfied that the ecosystem maps accurately represent IP dynamics in the CC-Sector innovation ecosystem and group consensus and theoretical saturation is reached (Bryman and Bell, 2015; Robson and McCartan, 2017).

In order to ensure internal validity of the NGT results, innovation ecosystem maps and expert panel feedback is correlated to empirical evidence, namely actual case examples of Type 1 New Entrants and their respective strategies to enter the CC-Sector, and theoretical concepts in the IP management and strategy literature. The following two subsections explain why we choose to apply visual ecosystem mapping to capture IP dynamics in the CC-Sector and describe the previously developed SVEL used in this process.

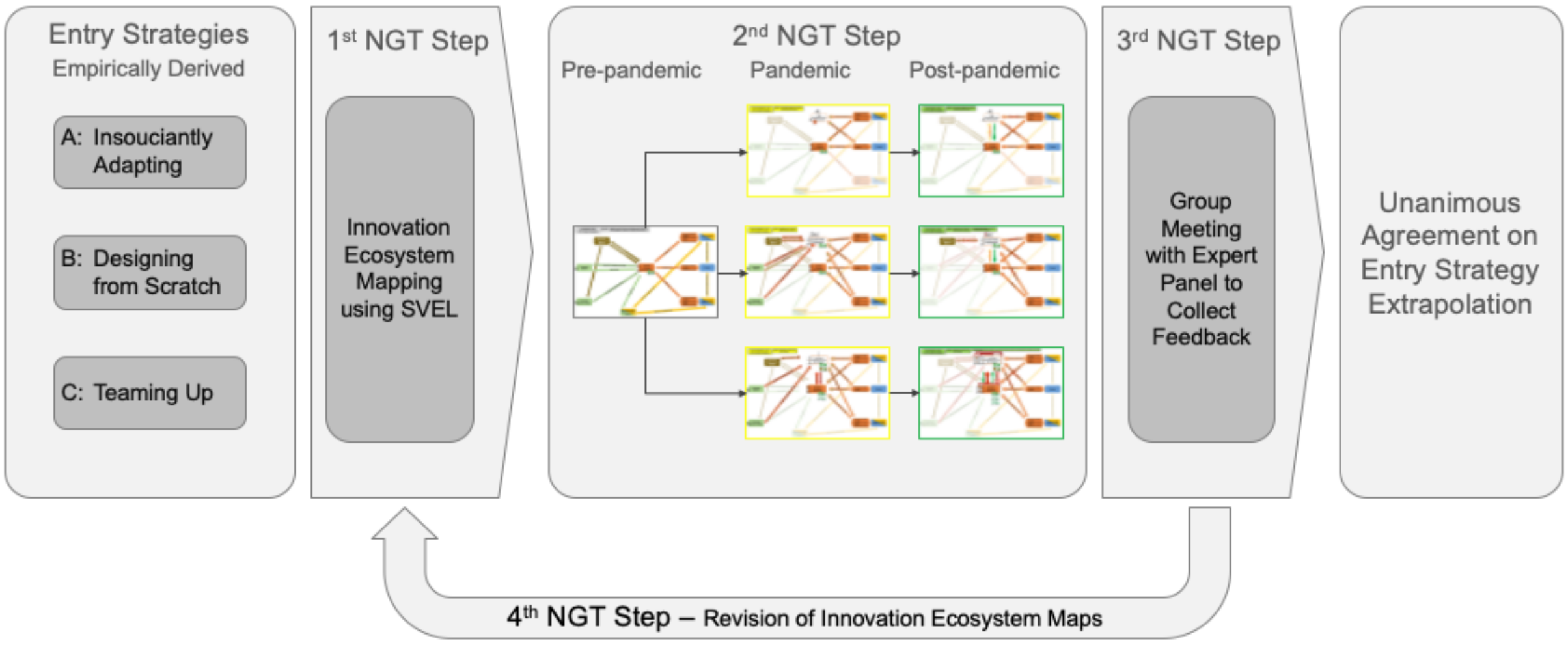

Figure 1 Overview of NGT-based research methodology

## 2.2 Innovation Ecosystem and Visual Mapping

We choose to represent the CC-Sector as an innovation ecosystem because it is highly appropriate for conceptually capturing dynamic processes involving multiple independent actors, artefacts and activities that evolve over time. The concept of the innovation ecosystem has developed from Moore’s (1993) analogy between the biological concepts of natural ecosystems and co-evolution of species, on one hand, and the organisation of business activities across industries and co-evolution of companies’ capabilities around new innovations, on the other hand. In a more recent innovation ecosystem conceptualisation, Granstrand and Holgersson (2020) established the link between an actor’s innovative performance and the evolving set of actors, activities and artefacts surrounding it, as well as the institutions and both complementary and substitute relations among them. Viewing the CC-Sector as an innovation ecosystem thus provides an effective conceptual basis for the identification of actors, the investigation of relationships among these actors, the localisation of innovation and hence ownership of resulting CC-IP and the usage of such CC-IP.

Moreover, Auerswald and Dani (2017) conceptually derived the adaptive cycle of economic ecosystems following their biological role models established by Holling (1992), in which an ecosystem’s inherent ability to absorb strong stochastic shocks and rapid disturbances by structural reorganisation is highlighted. Observing the CC-Sector through the innovation ecosystem lens thus also allows the observation of how actor roles, relationships and CC-IP ownership, as well as its usage, change dynamically as the Covid-19 pandemic unfolds, which is considered to represent the agent of disturbance (Holling, 1992) at this juncture.

Network visualisations have rarely been used in the context of the management sciences and ecosystem research, despite their ability to uncover information in empirical data that was previously not apparent (Basole, 2009), thereby offering opportunities for expanding the theoretical body of knowledge and gaining valuable insights for practitioners. Examples that nevertheless highlight the feasibility and benefits of ecosystem visualisations include Phillips and Srai's (2018) qualitative ecosystem mapping for the purpose of determining boundaries in emerging innovation ecosystems and Basole et al.'s (2018) web-based, easy-to-use and adaptable tool providing triangulated insight into temporal ecosystem dynamics for corporate executives. Furthermore, Urmetzer, Gill and Reed (2018) conducted ecosystem value mapping to visualise value creation and capture and to understand a case company's external complexity.

Due to the lack of standardisation in the emerging ecosystem visualisation methods, Moerchel et al. (2021) extended Urmetzer, Gill and Reed's ecosystem value mapping by developing a systematic approach to mapping structural elements, structural changes and dynamic forces and effects in business and innovation ecosystems, which they call the Standardised Visual Ecosystem Language. In this study, we take advantage of the benefits of a standardised ecosystem visualisation technique by applying this systematic approach for the purpose of understanding the IP-related risks and uncertainties in the CC-Sector during and after the Covid-19 pandemic induced crisis.

## 2.3 Standardised Visual Ecosystem Language

We adopt the novel Standardised Visual Ecosystem Language (SVEL) developed by Moerchel et al. (2021) to map the CC-Sector as an innovation ecosystem and capture the IP-related challenges, risks and uncertainties as the Covid-19 pandemic unfolds starting from the pre-pandemic phase through the pandemic to the post-pandemic phase. Based on the work of Urmetzer, Gill and Reed (2018), Moerchel et al. (2021) developed the SVEL to address the methodological gap of visually capturing and externally representing structural changes and their dynamic effects in an evolving ecosystem in a standardised fashion. This novel SVEL consists of three clusters of standardised external representations (see Table 3): (i) structural elements, (ii) structural changes and (iii) dynamic forces and effect.

### 2.3.1 Structural Elements

The formulation and representation of structural elements in the ecosystem was informed by Adner and Kapoor's (2010) generic ecosystem framework and to a larger extent by Granstrand and Holgersson's (2020) recent innovation ecosystem conceptualisation, which list actors, artefacts, and relationships among them as essential ingredients. These are effectively represented by the actor, goods flow and value proposition symbols in the SVEL, respectively, whereas colour coding is used to distinguish among actor roles. Particularly inspired by Adner and Kapoor (2010), SVEL is designed to focus only on those actors, artefacts and relationships that are essential for the formulation and delivery of a value proposition to customers, namely the focal firm, suppliers and complementors along with their respective products, component inputs and complementing offers. In addition, during SVEL development Moerchel et al. (2021) determined that ecosystem actors often occupy dual roles, such as the supplier-complementor or customer-complementor combinatorial roles.

### 2.3.2 Structural Changes

The external representations of ecosystem structural changes was inspired by Auerswald and Dani's (2017) notion of the ecosystem adaptive cycle and particularly the phases of release and reorganisation that are typically triggered by some form of ecosystem external disturbance. Moerchel et al. (2021) identified that these are best expressed by the evolution of actor roles, the diversion of or emergence of alternative goods flows, and the partial restriction or even full suspension of goods flows in an ecosystem.

### 2.3.3 Dynamic Forces and Effects.

These external representations were devised by Moerchel et al. (2021) in order to either capture effects on the relationship among actors that result from the ecosystem's evolution or to capture the external forces triggering phase changes in an ecosystem's adaptive life cycle (Auerswald and Dani, 2017). The standardised SVEL representations in the dynamic forces and effects cluster contain exertion of leverage or power by one ecosystem actor on another, the growth and shrinkage of goods flows, and changes in value capture by one ecosystem actor from the relationship with another actor.

## 3.0 Results and Discussion

In this section we present our results from the iterative NGT procedure, namely the empirically informed SVEL extension and adaptation and the CC-Sector innovation ecosystem visualisations for each of the three entry strategies during the empirically analysed pre-pandemic and pandemic phase of the Covid-19 induced crisis, as well as the conceptually extrapolated post-pandemic phase. Both the adapted and extended SVEL and the CC-Sector innovation ecosystem maps were generated iteratively and concurrently during the NGT structured process. The SVEL adoption and extension is presented first and followed by a description and discussion of the entry strategy scenarios and respective innovation ecosystem maps.

### 3.1 SVEL Adaptation and Extension for Innovation Ecosystem Visualisation

Moerchel et al. (2021) developed the SVEL for the purposes of visualising structural elements in an ecosystem, as well as capturing structural changes and dynamic effects and forces as the ecosystem evolves throughout its adaptive cycle. Starting with the first iteration of the NGT procedure it became clear, however, that SVEL needed both an extension with new external representations and an adaptation of the existing external representations in order to address the research goals of this study, namely capturing IP-related dynamics, risks and uncertainties in the evolving CC-Sector innovation ecosystem. The following two subsections provide further details and include tables showing the adapted and extended external representations. We believe that our extensions and adaptations enhance the Moerchel et al.'s (2021) methodological contribution in the field of ecosystem visualisation by increasing its scope of application, versatility and generalisability

Table 3: Summary of the SVEL including external representations and nomenclature (Moerchel et al., Forthcoming)

| Category | External Representation | Nomenclature |
|---|---|---|
| Ecosystem Structural Elements | | Single Role Ecosystem Actor |
| | | Dual Role Ecosystem Actor |
| | | Triple Role Ecosystem Actor |
| | | Tangible Goods Flow<br>(direction from producing to consuming actor) |
| | | Intangible Goods Flow<br>(direction from producing to consuming actor) |
| | | Value Proposition<br>(offered by producing to consuming actor) |
| Ecosystem Actor Role Colour Coding | | Ecosystem Focal Firm |
| | | Customers |
| | | Suppliers |
| | | Complementors |
| | | Other Actors<br>(without tangible product / intangible service flow) |

| Category | External Representation | Nomenclature |
|---|---|---|
| Dynamic Forces Driving or Resulting from Ecosystem Structural Changes | | Exertion of Leverage / Power |
| | | Growth<br>(adapts to Colour Coding of respective Goods Flow) |
| | | Shrinkage<br>(adapts to Colour Coding of respective Goods Flow) |
| | | Increased Value Capture<br>(direction towards capturing Ecosystem Actor) |
| | | Decreased Value Capture<br>(direction towards capturing Ecosystem Actor) |
| Ecosystem Structural Changes | | Actor Role Change with Near Match to other Actor<br>(adapts to matching Ecosystem Actor Colour Coding) |
| | | Actor Role Change with Limited Match to other Actor<br>(adapts to matching Ecosystem Actor Colour Coding) |
| | | Goods Flow with Near Match to other Actor's Goods Flow<br>(adapts to matching Good Colour Coding) |
| | | Goods Flow with Limited Match to other Actor's Goods Flow<br>(adapts to matching Good Colour Coding) |
| | | Diversion of Goods Flow<br>(adapts to original Good Colour Coding) |
| | | Full Suspension of Product or Service Flow |
| | | Partial Restriction of Product or Service Flow |

#### 3.1.1 SVEL Extension - Mapping IP in ecosystem dynamics

As the primary purpose of this study is to delineate the IP-related challenges, risks and uncertainties in the evolving CC-Sector innovation ecosystem, Moerchel et al.'s (2021) SVEL needed to be extended by devising additional external representations for (i) the various types of CC-IP, (ii) IP ownership and usage, and most importantly (iii) dynamic development of IP during pandemic phase-changes. This CC-IP specific extension is shown in Table 4 and described in further detail below:

(i) *CC-IP Types.* In order to externally represent CC-IP in the innovation ecosystem, first and foremost a distinction between two forms of IP is introduced to the SVEL, namely formal CC-IP represented as octagons and informal CC-IP depicted as triangles (see top section of Table 4). While formal CC-IP is typically registered with governmental agencies and therefore public, legally protected and exclusive to the innovator, informal CC-IP tends to be unobservable by other actors and protected by alternative means (Bonakdar et al., 2017). Examples of formal CC-IP typically comprise patents, trademarks, design rights and copyrights, whereas informal CC-IP typically consists of trade-secrets, complexity and lead-time (Hall *et al.*, 2014). While Hall et. al (2014) uncovered empirically that firms prefer informal IP over formal IP to protect their innovations, Holgersson and Granstrand (2017) determined using surveys that technology firms seek formal protection, and patents in particular, to protect their innovation, prevent competitors' access and to secure a freedom-to-operate.

(ii) *CC-IP Ownership and access:* In order to visually capture the spatial developments of CC-IP in the innovation ecosystem, namely CC-IP ownership and access, the four edges of the ecosystem actor symbol were classified into five sections (as shown in the centre section of Table 4). CC-IP that is used by the ecosystem actor, but owned by another (3rd party), is located along the bottom edge of its actor symbol, whereas the colour-coded frames of the CC-IP symbols indicate whether access to this CC-IP is authorised via licensing (green), unauthorised due to lack of licensing (red) or publicly accessible due to an expired exclusive right, such as an expired patent (orange). The remaining three edges house CC-IP that is owned by the ecosystem actor to which it is attached and classified according to the temporal and spatial origin of the innovation underlying the respective CC-IP. This approach was a result of the close collaboration of the NGT expert panel and inspired by key concepts of IP management in open innovation (Granstrand and Holgersson, 2013, 2014). More specifically, the left edge of the ecosystem actor symbol houses background CC-IP , which is CC-IP developed and existing prior to the Covid-19 pandemic, while the right edge holds foreground CC-IP , which is CC-IP jointly developed in collaboration of two ecosystem actors and typically owned by and accessible to both actors. Finally, the top edge is home to two classifications of CC-IP, namely sideground CC-IP and paraground CC-IP, both of which denote CC-IP that is independently developed by the ecosystem actor. The major difference between these two CC-IP classifications, however, is that sideground CC-IP is developed by an ecosystem actor which has a collaboration agreement, such as a license, with another ecosystem actor, thereby ensuring authorised access to 3rd party's background CC-IP, on which the independently developed sideground CC-IP is based. Paraground CC-IP, on the other hand, is independently conceived by an ecosystem actor outside of any collaboration and, thus, potentially based on unauthorised access to 3rd party background CC-IP (background) if freedom-to-operate is not established. In addition, two arrows were added, one indicating the provision of access to CC-IP owned by one actor to another, and the other to highlight the transformation of CC-IP during pandemic phase transitions.

(iii) *CC-IP Dynamic Developments:* In order to map the dynamic developments of CC-IP in the innovation ecosystem as the phases change during the Covid-19 induced crisis, it turned out to be highly beneficial to equip each IP symbol on the map with spatial and temporal identifiers. The spatial identifier is unique to each IP symbol and helps to trace its origin to type of IP owner and collaboration project, whereas the temporal identifier counts the phases during which the IP existed, effectively measuring its age. The three example spatial identifiers ADP, DES, and COL in Table 4 represent the entry strategy in which the respective Type 1 New Entrants conceived the new CC-IP, namely (A) Insouciantly Adopting, (B) Designing from Scratch and (C) Teaming up, respectively. The three temporal identifiers used are T0, T+1 and T+2 to represent the pre-, during and post-pandemic phases during which incumbent firms' background CC-IP for example existed, respectively.

### 3.1.2 SVEL Adaptations – Proof of its Versatility

The majority of SVEL adaptations consist of customisations of the nomenclature for the purposes of the CC-Sector innovation ecosystem. More specifically, the ecosystem structural element called "Ecosystem Focal Firm" in the SVEL by Moerchel et al. (2021) is renamed to "Incumbent CC-Products Manufacturer" and the "Tangible Goods Flow" is renamed as "CC-Product". Thus, we initially adopt the perspective of the incumbent firm in the pre-pandemic phase, which is sensible due to the fact that the incumbent effectively acts as the integrator of component inputs to the CC-Product and is the owner of the relevant CC-IP during this phase. However, as Type 1 New Entrants enter the CC-Sector innovation ecosystem in the pandemic phase, our perspective also includes that of the new entrant in addition to the incumbent. This effectively leads to the adaptation of the ecosystem structural change element called "Actor Role Change" in the SVEL by Moerchel et al. (2021) to the respective "Type 1 New Entrant". Furthermore, other changes to SVEL's structural elements and structural changes were minor and consisted of adding a "Crisis-Critical" or "CC"-prefix in the respective nomenclature for suppliers and their component inputs, complementors and their complement offers, as well as matching goods flows and goods flow diversions. The nomenclature for the external representations of dynamic forces and effects remained unchanged. Table 5,6 and 7 below contain the SVEL adaptations devised by the NGT expert panel for the purposes of visualising the CC-Sector innovation ecosystem maps.

## 3.2 Visualisations and Discussion of the CC-Sector Innovation Ecosystem

We examine each of the 3 entry strategies empirically derived by Tietze et al. (2020) namely (A) insouciantly adopting, (B) designing from scratch and (C) teaming up, separately for each of the 3 pandemic phases, namely pre-pandemic, pandemic and post-pandemic.

Table 4: SVEL Extension with IP specific external representations

| Category | External Representation | Nomenclature | Description |
|---|---|---|---|
| Extension of Ecosystem Structural Elements for Special Intellectual Property Considerations in Crisis-Critical Sector | 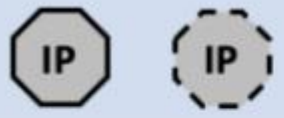 | Formal Intellectual Property | Intellectual Property that is public and exclusive to the innovator (usually for a limited time); **List of formal IP: patent, trademark, design right, copyright** (Bonakdar et al., 2017) (adapts to innovating Ecosystem Actor Colour Coding; dashed border represents complementary IP) |
| | 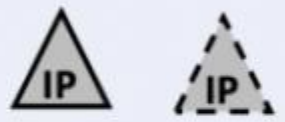 | Informal Intellectual Property | Intellectual Property that is not observable by other actors; **List of informal IP: trade secret, complexity, lead-time** (Bonakdar et al., 2017) (adapts to innovating Ecosystem Actor Colour Coding; dashed border represents complementary IP) |
| | 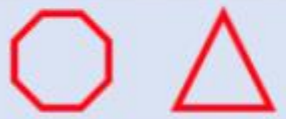 | Unauthorised IP Access | Intellectual Property used by the actor to which it is attached without any licensing agreements in place with the legal owner of the Intellectual Property |
| | 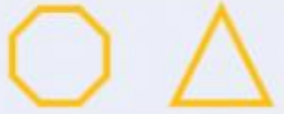 | Public Access to Expired IP | Intellectual Property used by the actor to which it is attached and for which the exclusive rights of the legal owner have expired or underlying knowledge advantage has become public |
| | 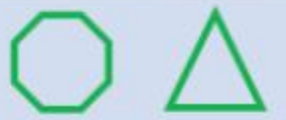 | Authorised IP Access | Intellectual Property used by the actor to which it is attached under a license agreement with the legal owner of the Intellectual Property |
| | 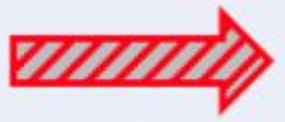 | CC-IP Access | Provision of access to CC-IP by the Ecosystem Actor owning the CC-IP to another Ecosystem Actor through licensing arrangements (adapts shaded Colour Coding of origin Ecosystem Actor) |

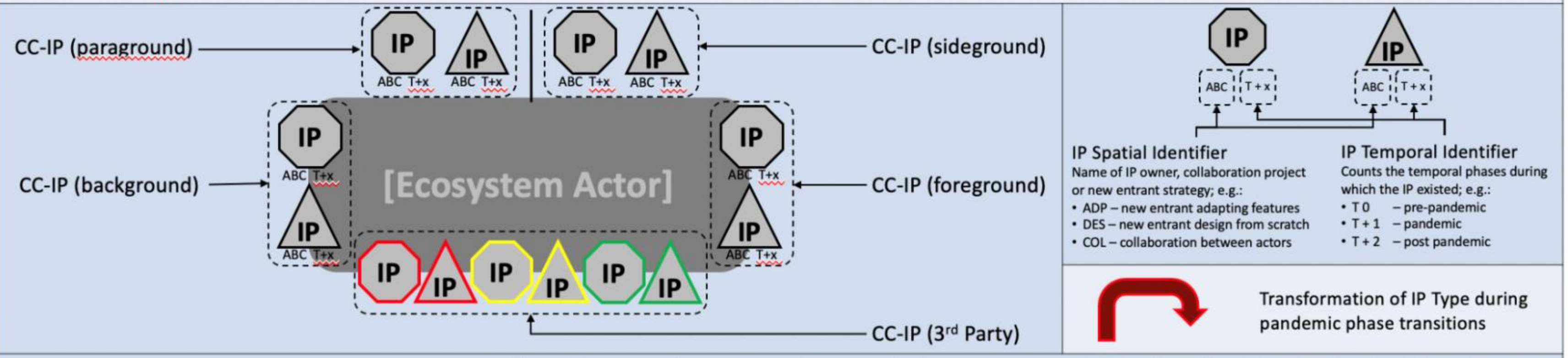

| | |
|---|---|
| CC-IP (background) (Granstrand & Holgersson, 2014) | Intellectual Property relevant or complementary for the manufacturing of CC-Products that is developed and existing prior to the Covid-19 Pandemic and owned by the incumbent or new entrant, respectively; complementary background IP has dashed border |
| CC-IP (foreground) (Granstrand & Holgersson, 2014) | Intellectual Property relevant for the manufacturing of CC-Products jointly developed in collaboration context during the Covid-19 Pandemic and owned by both the Incumbent CC-Product Manufacturer and New Entrant |
| CC-IP (sideground) (Granstrand & Holgersson, 2014) | Intellectual Property relevant for the manufacturing of CC-Products individually developed in a collaboration context during the Covid-19 Pandemic and owned by either the Incumbent CC-Product Manufacturer or New Entrant |
| CC-IP (paraground) (newly defined) | Intellectual Property relevant for the manufacturing of CC-Products individually developed by a New Entrant outside of a collaboration context during the Covid-19 Pandemic and potentially based on 3rd party background CC-IP |
| CC-IP (3rd Party) | Crisis-Critical Intellectual Property used by the Ecosystem Actor to which it is attached for the manufacture of CC-Products, but developed and owned by another Ecosystem Actor |

Table 5: Adaptation of SVEL Structural Elements

| Category | External Representation | Nomenclature | Description |
|---|---|---|---|
| Adaptation of Ecosystem Structural Elements for the Crisis-Critical Sector | 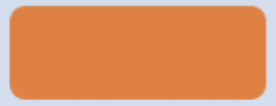 | Incumbent CC-Products Manufacturers | Firms that have been manufacturing Crisis-Critical Products (CC-Products) before firms from non-Crisis-Critical Sectors rush into Crisis-Critical Sectors to support meeting the upsurge in demand caused by the global Covid-19 Pandemic<br>Assumption: a market equilibrium exists among Incumbent CC-Product Manufacturers, such that innovation and hence competitive dynamics among them can be ignored |
| | 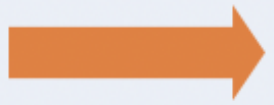 | CC-Product Flows from Incumbent Firm | Crisis-Critical Products manufactured and offered by Incumbent CC-Products Manufacturers (e.g. diagnostic tests, treatments, ventilators, vaccines, etc.) |
| | 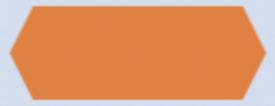 | Value Proposition of Incumbent Firms' CC-Products | Value proposition of Incumbent's CC-Products to Customers / End-Users |
| | 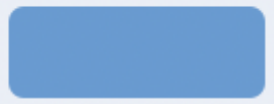 | Customers / End-Users | Firms, governmental / non-governmental organisations and individuals requiring CC-Products for the prevention, diagnosis / treatment of Covid-19 (e.g. hospitals/doctors, patients, NHS, etc.) |
| | 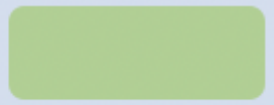 | CC-Suppliers | Firms that typically supply CC-Product and/or service components to Incumbent CC-Products Manufacturer for the integration/bundling into its CC-Products |
| | 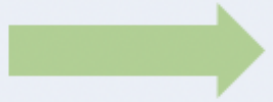 | Tangible CC-Component Input | Products supplied by CC-Suppliers to Incumbent CC-Products Manufacturer for integration/bundling into its CC-Products (e.g. sub-assemblies, raw materials, consumables, etc.) |
| | 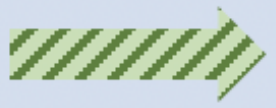 | Intangible CC-Component Input | Services supplied by CC-Suppliers to Incumbent CC-Products Manufacturer for integration/bundling into its CC-Products (e.g. lab services, software, background-IP etc.) |
| | 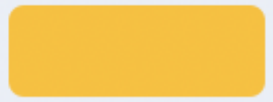 | CC-Complementors | Firms that typically provide other CC-Product and/or service complements for downstream integration/bundling by Customers / End-Users such that the combination allows the prevention, diagnosis or treatment of Covid-19 |
| | 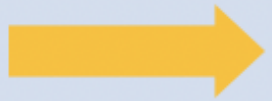 | Tangible CC-Complement Offers | Products offered by CC-Complementors directly for downstream integration/bundling by Customers / End-Users such that the combination allows the prevention, diagnosis and treatment of Covid-19 (e.g. consumables, etc.) |
| | 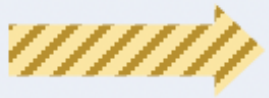 | Intangible CC-Complement Offers | Services offered by CC-Complementors directly for downstream integration/bundling by Customers / End-Users such that the combination allows the prevention, diagnosis and treatment of Covid-19 (e.g. lab services, software, etc.) |
| | 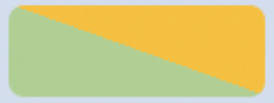 | CC-Supplier/CC-Complementor (Dual Role) | Firms occupying a dual role in the CC-Sector ecosystem by typically supplying products and/or services both as CC-Components to Incumbent CC-Product Manufacturers and as CC-Complements to Customers / End-Users |
| | 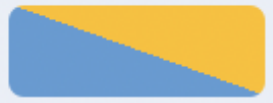 | CC-Customer/CC-Complementor (Dual Role) | Firms, organisations and individuals occupying a dual role in the CC-Sector ecosystem by typically purchasing CC-Products from Incumbent Manufacturer and supplying product and/or service complements to Customers / End-Users |
| | 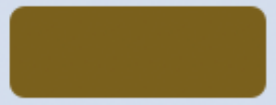 | Government | Governmental / Intergovernmental Agencies having oversight of CC-Product Approval, Certification or Registry (e.g. US Food and Drug Administration, European Patent Office) |
| | 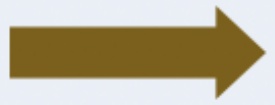 | Government-issued Approval / Certification / Registration of CC-Product | Approval, Certification or Registry of CC-Products by Government |

Table 6: Adaptation of SVEL Structural Changes

| Category | External Representation | Nomenclature | Description |
|---|---|---|---|
| Adaptation of Ecosystem Structural Changes in Crisis-Critical Sector | 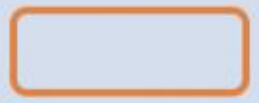 | Type 1 New Entrant – Large Manufacturers and Technology Giants | Large Manufacturers with spare valuable resources, tech giants with diverse capabilities and rich resources or other firms with relevant expertise, resources and competence either repurposing their manufacturing capacities to cover fixed costs and preserve their workforce or diversifying into the Crisis-Critical Sector from non-Crisis-Critical Sectors in order to support meeting the upsurge in demand for CC-Products caused by the Covid-19 Pandemic |
| | 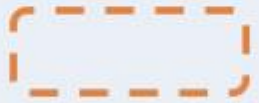 | Type 1 New Entrant – Agile Manufacturers | Agile manufacturers with particular and complementary skills entering the Crisis-Critical Sector in order to support local / regional effort to mitigate supply chain bottleneck for CC-Products and alleviate crisis situation caused by the Covid-19 Pandemic |
| | 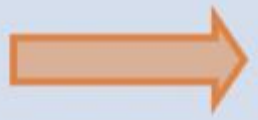 | Enhanced CC-Product | CC-Product manufactured by New Entrant (typically Large Manufacturers or Tech Giants) either designing from scratch or teaming up with Incumbent CC-Product Manufacturer; access to background CC-IP may or may not be granted by Incumbent Manufacturer; CC-Product may be considerably enhanced compared to the Incumbents' either from a product feature or manufacturing process perspective |
| | 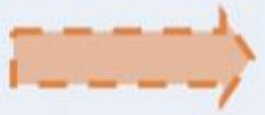 | Limited Matching CC-Product | CC-Product manufactured by New Entrant (typically Agile Manufacturers) insouciantly adapting certain features of Incumbent Manufacturer's CC-Product without conducting freedom-to-operate analysis or licensing background CC-IP; this CC-Product lacks particular features due to resource constraints, but satisfies requirements for being minimally clinically acceptable |
| | 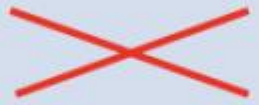 | Full Supply Suspension | The supply of tangible products or intangible services is suspended completely between two Crisis-Critical ecosystem actors |
| | 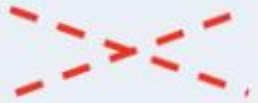 | Partial Supply Restriction | The supply of tangible products or intangible services is limited or restricted between two Crisis-Critical ecosystem actors (e.g. provision only under licensing contracts, limitation of quantities, restrictions on downstream use/application, etc.) |
| | 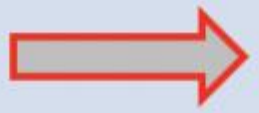 | CC-Component Input Diversion | Supply of Tangible and Intangible CC-Component Inputs by CC-Suppliers to Incumbent CC-Product Manufacturer diverted to New Entrants for the manufacture of Matching CC-Products (adapts to original Good Colour Coding) |

Table 7: Adaptation of SVEL Dynamic Effects

| Category | External Representation | Nomenclature | Description |
|---|---|---|---|
| Adaptation of Dynamic Forces Driving or Resulting from Ecosystem Structural Changes in the Crisis-Critical Sector | 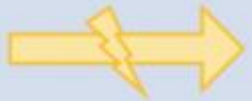 | Exertion of Leverage / Power | Exertion of bargaining power of CC-Suppliers and/or CC-Complementors, exertion of bargaining power of buyers or exertion of executive power of Government |
| | 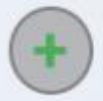 | Sales Growth | Increase in sales numbers of the Incumbent's or New Entrants' CC-Product (matching or imitating) relative to the previous phase (i.e. pre-, during or post-pandemic); (adapts to Colour Coding of respective Goods Flow) |
| | 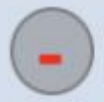 | Sales Shrinkage | Decrease in sales numbers of the Incumbent's or New Entrants' CC-Product (matching or emulating) relative to the previous phase (i.e. pre-, during or post-pandemic); (adapts to Colour Coding of respective Goods Flow) |
| | 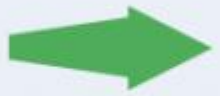 | Increased Value Capture | Increasing trend in value captured by the receiving actor either economically (e.g. higher price for its goods supply, licensing fees, etc.) or through access to improved / enhanced products and services |
| | 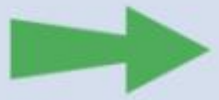 | Decreased Value Capture | Decreasing trend in value captured by the receiving actor either economically (e.g. lower price for its goods supply due to increasing competition and lack of differentiation) or through loss of access to products and services |

Table 8 supplements these three entry strategies, which were deduced by Tietze et al. (2020) and previously aligned to Type 1 New Entrant groups, with empirically identified motives to enter the CC-Sector innovation ecosystem and expected behaviour in the post-pandemic phase. While the pandemic and post-pandemic phases are different for each of the entry strategy scenarios, they all start from the same initial condition in the pre-pandemic phase, meaning that the latter is considered only once. Furthermore, the phase changes are determined on the basis of the crisis definition by Tietze et al. (2020), which stipulates that both a threat to human species needs to exist and risk criticality needs to be elevated by an applicable world body [4].

Table 8: Motives to enter CC-Sector and expected behaviour in post-pandemic phase for each alignment of entry strategy and Type 1 New Entrant group

| **Entry Strategy** | **Type 1 New Entrant group** | **New Entrant motives** | **Expected behaviour of New Entrant in post-pandemic phase** |
|---|---|---|---|
| (A) Insouciantly Adopting | (4) Agile manufacturers with particular and complementary skills to the CC-Sector | • Support and give-back to local/regional community<br>• Support local/regional effort to mitigate supply chain bottleneck for CC-Products and alleviate crisis situation, also known as the citizen supply chain (Larrañeta, Dominguez-Robles and Lamprou, 2020) | ➔ Likely to exit CC-Sector after crisis situation is terminated |
| (B) Designing from Scratch | (3) Firms with relevant expertise, resources and competence | • Entrepreneurship: motivated by confidence that relevant expertise leads to substitute/improved/ competitive CC-Products<br>• For complex CC-Products (e.g. ventilator), independent CC-Product development costs are high (Bottomley, 2020) and time to achieve satisfactory return-on-invest may extend beyond the termination of the crisis | ➔ Likely to remain in CC-Sector after crisis situation is terminated |
| (C) Teaming Up | (1) Manufacturers with spare, valuable resources and capabilities<br>(2) Tech giants with diverse capabilities and rich resources | • Utilisation of idle manufacturing capacities to cover fixed costs and to preserve workforce though crisis (López-Gómez *et al.*, 2020)<br>• Motivated to recover sunk costs in repurposing manufacturing<br>• Create positive image in the public | ➔ Likely to remain in CC-Sector until non-CC-Sector picks up to pre-pandemic levels |

[4] Accordingly, the phase change from pre-pandemic to pandemic, or the onset of the Covid-19 pandemic induced crisis, is determined by the WHO Director General's declaration of Covid-19 as a pandemic on 11 March 2020 (World Health Organisation, 2020c), which followed a dramatic surge in demand for CC-Products worldwide. The phase change from pandemic to post-pandemic, or the termination of the Covid-19 pandemic induced crisis, will probably be heralded by a continuous drop in demand for CC-Products due to decreasing numbers of patients and officially declared by the WHO in a similar fashion as the containment of the SARS outbreak in 2003 (World Health Organisation, 2003).

Before mapping the CC-Sector innovation ecosystem for each of the entry strategy scenarios and during each of the pandemic phases, we found it to be useful to synthesise archetypical intentions of new entrants to stay in the post-pandemic CC-Sector because it provides guidance on strategic behaviour as the Covid-19 induced crisis advances through the pandemic into the post-pandemic phase. As shown in Table 8, group 4 Type 1 New Entrants insouciantly adopting incumbent firms' CC-Products form a citizen supply chain (Larrañeta, Dominguez-Robles and Lamprou, 2020) with a focus on supporting the local/regional effort to alleviate CC-Product supply chain bottlenecks during the crisis situation and are therefore likely to exit the CC-Sector after the crisis situation is terminated.

Furthermore, group 3 Type 1 New Entrants designing CC-Products from scratch follow an entrepreneurial approach and have a tendency towards staying in the CC-Sector after the crisis situation is terminated in order to recover their substantial CC-Product development cost (Bottomley, 2020). Finally, group 1 and 2 Type 1 New Entrants teaming up with incumbent firms are concerned with retaining utilisation of their idle manufacturing capacities by repurposing (López-Gómez *et al.*, 2020) and are hence likely to remain in the CC-Sector as long as demand for their non-CC-Products remains below pre-pandemic levels.

In the following subsections, we describe and discuss the CC-Sector innovation ecosystem maps that result from applying the adapted and extended SVEL to each of the three Type 1 New Entrant group to entry strategy alignments during the iterative NGT procedure. We find these to be very helpful for deriving an understanding of IP-related risks, uncertainties and challenges from both the incumbent firms' and Type 1 New Entrant group's perspectives.

### 3.2.1 Pre-pandemic View of CC-Sector Innovation Ecosystem

Figure 2 depicts the CC-Sector innovation ecosystem using the adapted and extended SVEL before the Covid-19 pandemic started to unfold, including the relevant actors, activities, artefacts and relations among them in line with Granstrand and Holgersson's (2020) definition of the innovation ecosystem. The Incumbent CC-Product Manufacturers (representing a group of such firms) are positioned as focal firms at the centre of the CC-Sector innovation ecosystem because they are bundling all necessary components into the CC-Products (Adner and Kapoor, 2010). The CC-Products, such as PPE, ventilators or drugs (Tietze *et al.*, 2020), are, in turn, offered by the Incumbent CC-Product Manufacturer as a customised value proposition to diverse customer groups, namely Patients, Publicly Funded Healthcare Systems or Private Hospitals and Doctors [5].

During the process of CC-Product development in the pre-pandemic phase, incumbent firms endogenously develop innovations and thus intellectual assets covering both product specification and manufacturing process information, for which they seek legal protection thus building a portfolio of formal and informal IP. More specifically, as part of this CC-Product development process, Governmental and Intergovernmental Agencies, such as the Federal

[5] While Patients primarily purchase PPE for the individual prevention of contracting the disease (World Health Organisation, 2020a) and drugs for the treatment of milder symptoms caused by Covid-19, Publicly Funded Healthcare Systems and Private Hospitals and Doctors purchase drugs, ventilators and PPE from Incumbent CC-Product Manufacturers both to treat Patients in critical condition who require intensive care and to protect their staff when diagnosing and treating infected persons with severe symptoms (Phua *et al.*, 2020), respectively.

Drug Administration, grant Certifications and Approvals following clinical trials of CC-Products that are common and necessary in the CC-Sector (López-Gómez *et al.*, 2020).

Furthermore, other Government Agencies, such as the European Patent Office, issue formal IP rights, such as patents, registered designs and trademarks (Nicholson Price II, Rai and Minssen, 2020), thus enabling incumbent firms to exploit their innovations and earning a return on their investments in CC-Product related research and development (Contreras *et al.*, 2020). When moving on into the pandemic phase, this IP will form the background Crisis-Critical IP, namely CC-IP (background), that is owned by the Incumbent CC-Product Manufacturer since the development of the underlying innovation prior to the start of a Covid-19 induced crisis (Tietze *et al.*, 2020).

Furthermore, Incumbent CC-Product Manufacturers rely on tangible and intangible component inputs from Component Suppliers, Universities or Research Institutes and other Manufacturers with CC-Product Relevant Expertise in order to deliver integrated and bundled CC-Products and offer customised value propositions to various customer groups. Tangible component inputs typically include CC-Product Subassemblies manufactured by Component Suppliers, such as air compressors, control valves and electronic control circuitry typically found in mechanical ventilators (Chatburn, 1991). Intangible component inputs, on the other hand, can entail licenses to use formal and informal CC-IP (background), such as specialised manufacturing methods for biologic products previously developed and owned by University or Research Institute laboratories (Nicholson Price II, Rai and Minssen, 2020). Furthermore, as CC-Product complexity increases, the supplier landscape is likely to become more diverse in both component inputs and number of suppliers (Hobday, Davies and Prencipe, 2005).

Another essential concept in the innovation ecosystem paradigm is complementarity both as a source of additional value of the focal offering (Adner and Kapoor, 2010) and as the counterpart relationship to competition (Granstrand and Holgersson, 2020). In the pre-pandemic CC-Sector innovation ecosystem this concept is expressed by crisis-critical offerings that are supplied by complementors directly to customer groups for downstream bundling with incumbent firms' CC-Products, thus enhancing the combined value proposition. More specifically, other Manufacturers with CC-Product Relevant Expertise offer Intensive Care Equipment and PPE directly to Private Hospitals and Doctors, which require these complementing offers in order to provide Intensive Care and Special Treatment using incumbent firms' CC-Products, such as Ventilators and Covid-19 specific Drugs, effectively and safely (World Health Organisation, 2020a). The CC-Sector innovation ecosystem visualisation in Figure 2 also shows that the concept of complementarity in the CC-Sector means that some actors occupy dual roles, namely either both as supplier and complementor as in the case of Manufacturers with CC-Product Relevant Expertise or both as customer and complementor as in the case of Private Hospitals or Doctors.

Finally, the pre-pandemic CC-Sector innovation ecosystem shown in Figure 2 is assumed to be in an equilibrium state that is historically resilient to small displacements in demand for CC-Products. In the context of comparative statics, such an equilibrium state is said to exhibit stability of the first kind in the small (Samuelson, 1941).

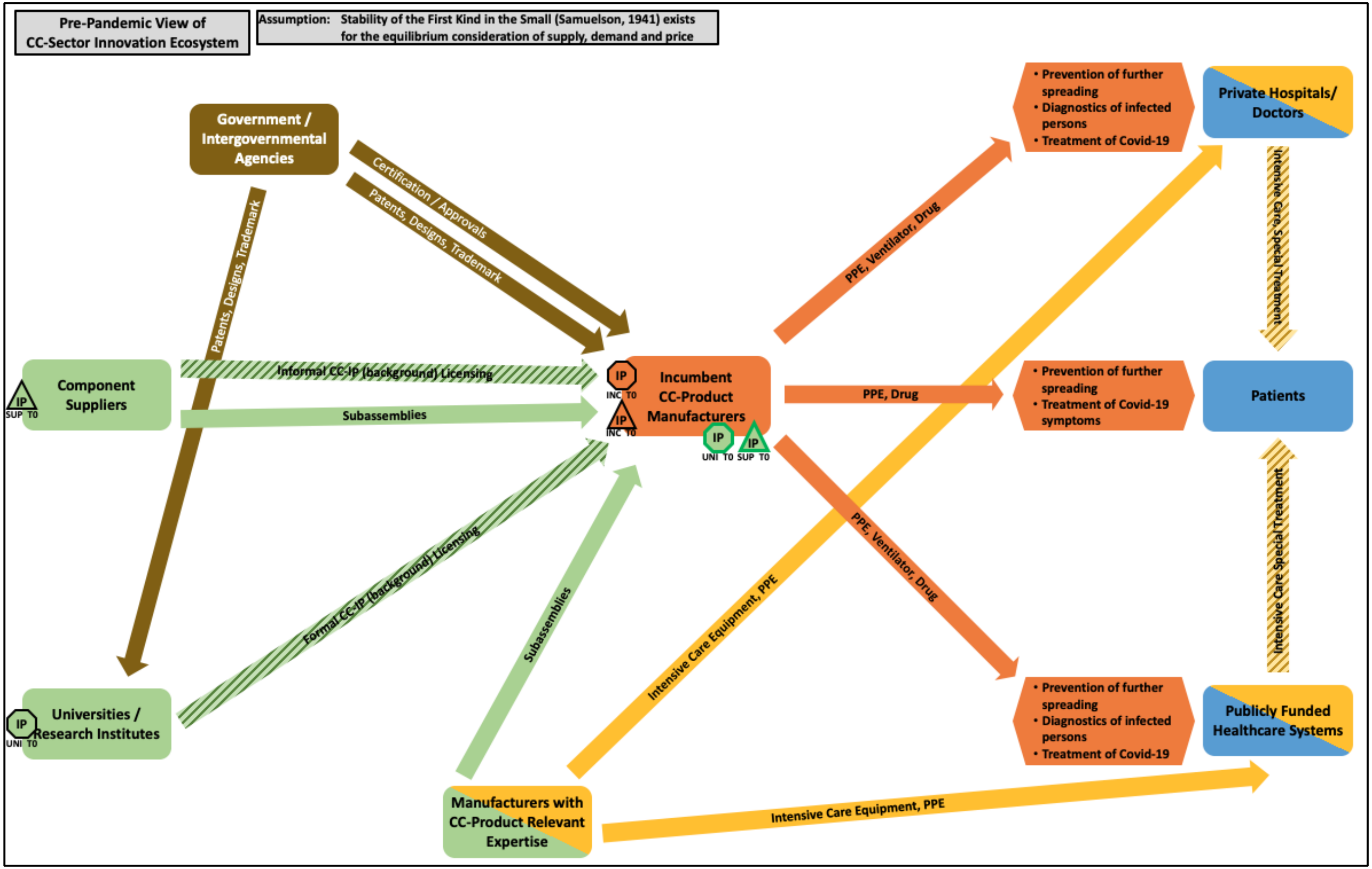

Figure 2: Visualisation of pre-pandemic CC-Sector Innovation Ecosystem

The Covid-19 pandemic induced crisis situation, however, manifests itself in a sudden and unprecedented surge in demand for the CC-Products (Contreras *et al.*, 2020; Phua *et al.*, 2020) representing a stochastic demand shock that forces the CC-Sector innovation ecosystem to enter into a release and reorganisation phase of its adaptive lifecycle (Auerswald and Dani, 2017). This evolution of the CC-Sector innovation ecosystem is further analysed in the following subsections which describe and discuss the application of the adapted and extended SVEL to visualise the pandemic and post-pandemic phases for each of the three entry strategies.

### 3.2.2 Entry Strategy A – Insouciantly Adopting

Entry strategy A is characterised by Type 1 New Entrants designing and manufacturing a CC-Product while insouciantly adopting features of incumbent's CC-Products through copying or reverse engineering without clearing IP that might be owned by incumbent firms, such as by conducting a freedom-to-operate analysis (Tietze *et al.*, 2020). Type 1 New Entrants following this strategy typically are agile manufacturers with particular and complementary skills to the CC-Sector, which are herein denoted as group 4 Type 1 New Entrants [6]. By not conducting due diligence and thereby accepting the risk of infringing formal CC-IP (background) that is owned by the Incumbent CC-Product Manufacturer, these Type 1 New Entrants may face infringement claims by the incumbent or injunctions against their CC-Product, even though incumbents might not exercise their rights due to potential reputational damages during the pandemic or because they are simply busy with scaling up production (Tietze *et al.*, 2020).

As shown in Figure 3 visualising entry strategy A in the pandemic CC-Sector innovation ecosystem, Type 1 New Entrants from group 4 tend to only adopt essential features of incumbent firms' CC-Product due to their relative small size and limited resources (Tietze *et al.*, 2020), nevertheless meeting minimally clinically acceptable specifications (López-Gómez *et al.*, 2020). This minimal viable CC-Product is then offered to Patients, as well as Private Hospitals and Doctors, concurrently to the original CC-Product by incumbent manufacturers in order to alleviate the surge in demand during the pandemic phase of the crisis on a local or regional scale (Manero *et al.*, 2020).

During the process of developing and manufacturing a minimal viable CC-Product, Type 1 New Entrants from group 4 typically focus on essential functionalities, thus omitting more advanced features of incumbent's CC-Products. More specifically, the urgency and volume of demand dictated by the Covid-19 pandemic induced crisis (Contreras *et al.*, 2020; Phua *et al.*, 2020) coupled with endogenous limitations in manufacturing resources and capabilities (Manero *et al.*, 2020; Tietze *et al.*, 2020) require firms in this Type 1 New Entrant group to realise substantial lead-time and cost reductions and to focus on an optimised performance level for the local and regional customer groups [7]. Group 4 Type 1 New Entrants essentially satisfy the criteria for conducting frugal innovation (Weyrauch and Herstatt, 2017), thereby introducing micro level technological discontinuities (Garcia and Calantone, 2002) that are likely to be new to incumbent firms. These frugal innovations materialise as newly developed informal CC-IP that is based on both group 4 firms' complementary background IP and incumbent firms' CC-IP (background). As exemplified by the coalition of 3D printer manufacturers (Larrañeta, Dominguez-Robles and Lamprou, 2020; Manero *et al.*, 2020), this newly developed informal CC-IP could relate to the use of substitute materials that are available to group 4 firms from their existing suppliers, thus realising lead-time and cost-reductions, and are necessary in order to manufacture the minimal viable CC-Products with equipment and process technology available to them.

[6] Examples are Formula 1 teams joining forces to launch Project Pitlane and reverse-engineering and rapidly manufacturing a continuous positive airway pressure device (Richards, 2020), luxury brand manufacturers, such as LVMH, L'Oreal and Estee Lauder, switching their production facilities to making hand sanitisers (Pays, 2020), cloth manufacturers, such as Parkdale Mills, repurposing production lines to making PPE (Ma, 2020), and 3D printer manufacturers, such as Prusa3D, Formlabs and Stratasys, enabling their widely distributed customer base to produce standardised and validated PPE designs for their respective communities (Manero *et al.*, 2020).

As this novel informal CC-IP is developed by group 4 Type 1 New Entrants neither in collaboration with the Incumbent CC-Product Manufacturers nor on the basis of authorised access to the incumbent firms' CC-IP (background), it does not qualify as foreground or sideground IP (Granstrand and Holgersson, 2014). It is thus herein newly defined as paraground CC-IP. While group 4 firms typically fail to conduct a freedom-to-operate analysis prior to insouciantly adopting features in its minimal viable CC-Product, to which incumbent firms have an exclusive right through formal ownership of CC-IP (background), the Incumbent CC-Product Manufacturer tends to avoid legal action against group 4 Type 1 New Entrants during the pandemic because of potential reputational damage (Tietze *et al.*, 2020) and due to demand for CC-Products anyway exceeding the incumbent firms' production capacities (Ambrosio, 2020). Nevertheless, this novel CC-IP (paraground) developed by group 4 firms following this entry strategy is potentially still of interest to incumbent firms. After the Covid-19 pandemic terminates and the demand for CC-Products decreases towards pre-pandemic levels and below incumbent firms' production capacity, as shown in Figure 4, the Incumbent CC-Product Manufacturer may however choose to respond to group 4 Type 1 New Entrants, for example, by filing infringement claims against them for unauthorised usage of its formal CC-IP (background) [8], to which the latter gained access through copying and reverse engineering (Tietze *et al.*, 2020). Without having conducted a freedom-to-operate analysis prior to entering the CC-Sector with its minimal viable CC-Product, group 4 Type 1 New Entrants now face the risk of having to compensate the incumbent for lost profits or at least a reasonable royalty (Poltorak and Lerner, 2011, p. 120). Furthermore, infringement claims by incumbent firms could also aim for an injunction against the group 4 Type 1 New Entrant barring any further infringement (Poltorak and Lerner, 2011, p. 123), thus fully suspending the supply of minimal viable CC-Products to Patients and Hospitals or Doctors. This post-pandemic scenario and the IP-related dynamics are captured in Figure 4.

[7] For instance, Manero et al. (2020) reported that a coalition of 3D printer manufacturers in the U.S. along with their international customer base succeeded in optimising face shield visor designs for distributed additive manufacturing using 3D printers. Designs applicable for 3D printing typically consist of adapted shapes and raw material specifications that realise short manufacturing lead-times and material waste reduction (Larrañeta, Dominguez-Robles and Lamprou, 2020). Despite inferior material properties, such as higher porosity, and higher variability in the production process, 3D printed visors manufactured by the coalition were cleared for use under the Federal Drug Administration's Emergency Use Authorization in the U.S. (Manero *et al.*, 2020).

[8] A precedent was set by Allele Pharmaceuticals Inc., who sued Regeneron in White Plains, New York, and filed complaints against Pfizer Inc. and BioNTech SE in a federal court in San Diego, California, for applying its patented fluorescent protein called mNeonGreen while developing treatments of Covid-19 without a license (Yasiejko, 2020).

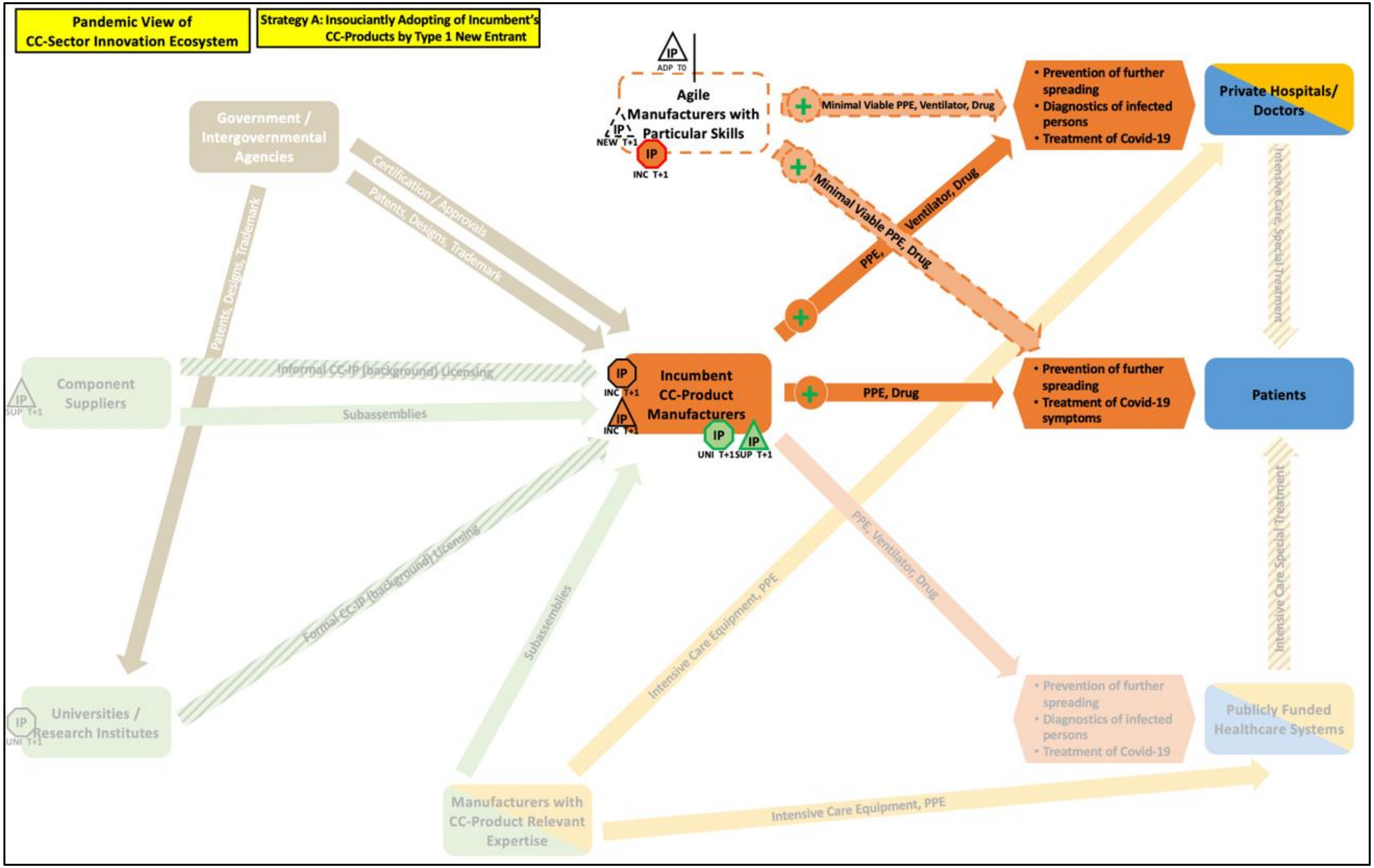

Figure 3: Visualisation of pandemic CC-Sector innovation ecosystem with group 4 Type 1 New Entrants insouciantly adopting incumbent firms' CC-product

Ultimately, Incumbent CC-Product Manufacturers need to decide whether the anticipated commercial benefit of compensation and injunction exceeds the cost of litigation (Poltorak and Lerner, 2011, p. 113f), which may not necessarily be the case with group 4 firms, namely Agile Manufacturers with Particular Skills due to their small scale operation (Tietze *et al.*, 2020). In addition, the incumbent may contemplate the strategic relevance of gaining access to the newly developed informal CC-IP (paraground) developed by group 4 Type 1 New Entrants during the pandemic phase.

As determined above, group 4 firms' informal CC-IP (paraground) could effectively represent frugal innovation bearing the potential for minimised resource requirements and cost reductions while still satisfying quality and performance (Tiwari and Herstatt, 2012). These characteristics may become relevant features of the CC-Product value chain in the 'new normal' of the post-pandemic phase due to an emerging general consensus on voluntary simplicity and affordability (Herstatt and Tiwari, 2020).

Thus, when deciding how to respond to the competitive threat posed by group 4 Type 1 New Entrants in the post-pandemic phase, Incumbent CC-Product Manufacturers need to reflect whether either the prospect of damage compensation by the infringing new entrant or the potential value of the frugal innovation captured in its CC-IP (paraground) outweigh the costs of litigation.

Group 4 Type 1 New Entrants, on the other hand, also face a strategic decision as the Covid-19 pandemic comes to an end and the CC-Sector transitions into the post-pandemic phase, namely whether to cease from the manufacturing of CC-Products or to continue, thus remaining in the CC-Sector. In the latter case, it would be prudent for the group 4 Type 1 New Entrants to negotiate licenses with the Incumbent CC-Product Manufacturer, thereby finally ensuring authorised access to its CC-IP (background). Offering the incumbent cross-licensing of its own CC-IP (paraground) may be a sensible approach by group 4 Type 1 New Entrants to build leverage, but incumbent firms would probably be inclined to pursue litigation in order to force the group 4 firms out from the contracting CC-Sector and nevertheless seek access to CC-IP (paraground) as part of the sought compensation.

In essence, considering the small scale of group 4 Type 1 New Entrants and their local to regional focus, it would be more realistic for these firms in this scenario to proactively cease manufacturing of CC-Products and follow and inside-out open innovation approach (Chesbrough, 2012) by out-licensing its CC-IP (paraground) to other actors in the CC-Sector, including incumbent firms. This strategic path avoids any IP infringement claims and potentially allows monetisation of its false-negative IP (Chesbrough, 2012).

### 3.2.3 Entry Strategy B – Designing from Scratch

When following entry strategy B, Type 1 New Entrants use their inherent technical competence and, if necessary, outside expert advice to design a CC-Product from scratch (Tietze *et al.*, 2020) with the goal to match the value proposition of incumbent firms' CC-Products.

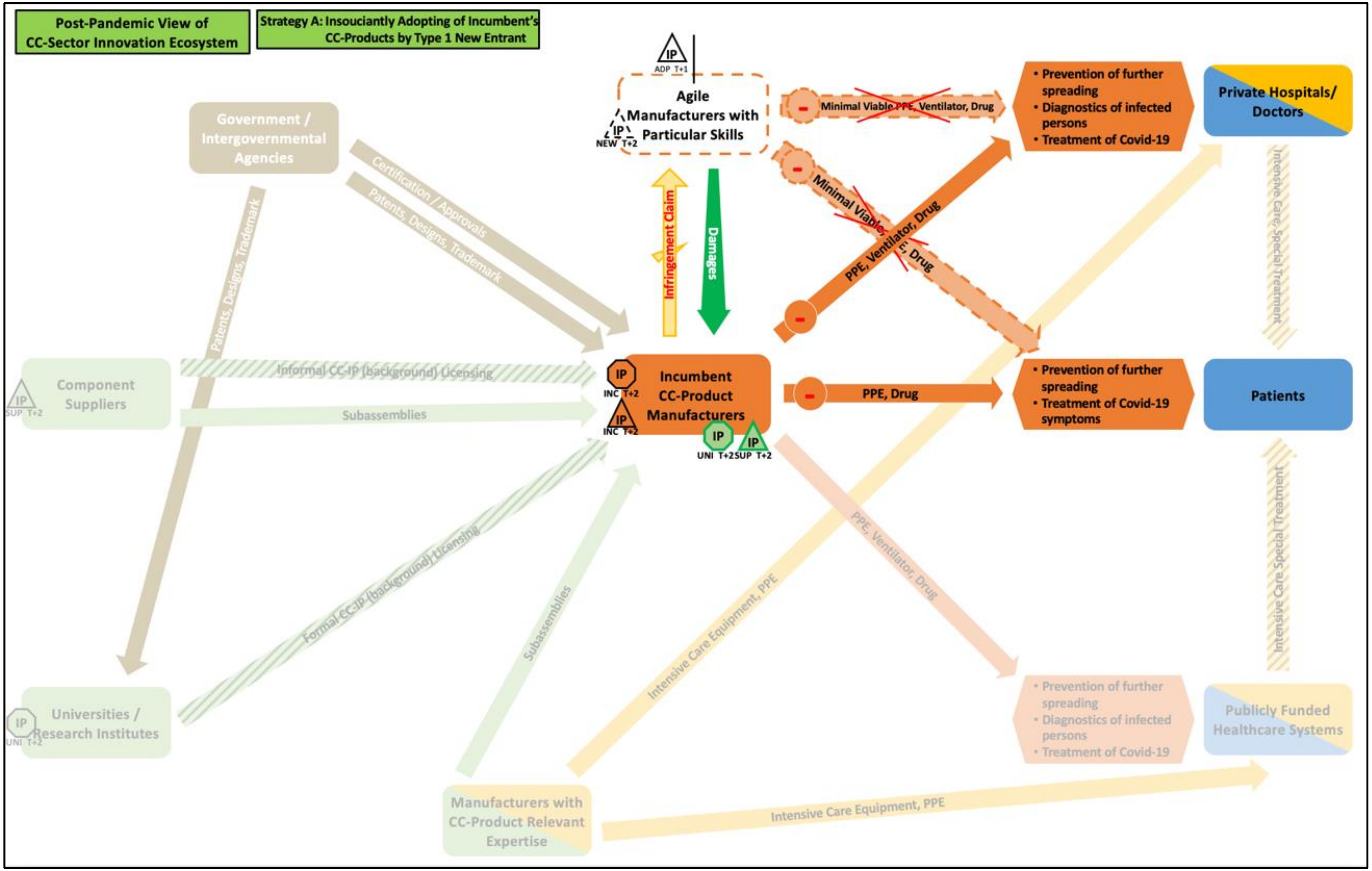

Figure 4: Visualisation of post-pandemic CC-Sector innovation ecosystem with Incumbent CC-Product Manufacturers filing an infringement claim against group 4 Type 1 New Entrant

These manufacturing firms entering the CC-Sector by newly developing CC-Products are herein denoted as group 3 Type 1 New Entrants[9]. Dictated by the dynamic upsurge in demand for CC-Products at the start of the pandemic, group 3 Type 1 New Entrants following entry strategy B may not have sufficient time to conduct thorough freedom-to-operate analyses, thus potentially infringing CC-IP (background) owned by Incumbent CC-Product Manufacturers (Tietze *et al.*, 2020) in a similar fashion as in the entry strategy A scenario. Group 3 Type 1 New Entrants choosing entry strategy B typically realise that they possess capabilities and resources to potentially design, develop and produce a substitute CC-Product with enhanced features that is equivalent to the incumbent firms' CC-Product, thereby representing an additional source during the pandemic phase when CC-Product demand substantially exceeds incumbent firms' manufacturing capacities.

Furthermore, being larger firms, Type 1 New Entrants of group 3 typically own formal and informal background IP that may be complementary, but new to the CC-Sector allowing them to introduce macro-level technological discontinuities to the sector (Garcia and Calantone, 2002). For example, Group 3 Type 1 New Entrants essentially have expertise and competences captured in complementary background IP to design and develop enhanced CC-Products from scratch[10], as well as the resources and scale to offer it to all customer groups in the CC-Sector innovation ecosystem, namely Patients, Publicly Funded Healthcare Systems and Private Hospitals or Doctors, on a national level, whose demand cannot be met by incumbent firms. Figure 5 below depicts group 3 Type 1 New Entrants designing CC-Products from scratch and introducing enhancements to the CC-Sector that are based on their complementary formal and informal background IP from outside the CC-Sector.

As group 3 Type 1 New Entrants design and develop CC-Products from scratch, thereby applying their complementary background IP to applications in the CC-Sector, as well as deviating from existing CC-Product designs in order to accommodate their existing manufacturing processes and equipment and supplier network, they potentially come up with innovative solutions to solve design problems or manufacturing challenges[11]. This does not only potentially lead to a CC-Product enhancement as discussed above, but also essentially represents novel CC-IP (Tietze *et al.*, 2020). The development of this novel CC-IP, however, comes at a cost as shown by the Dyson CoVent example, which reportedly required an investment of 20 million pounds (Bottomley, 2020). The magnitude of such investment combined with economic pressures during the crisis and entrepreneurial spirit dictate that group 3 Type 1 New Entrants seek options to realise a return on their investment (4iP Council, 2016). Next to seeking expedited regulatory approval of their enhanced CC-Products, which is typically a stringent requirement in the CC-Sector (López-Gómez *et al.*, 2020), it would be prudent for group 3 firms to also apply for formal protection of their newly developed CC-IP by registering patents, designs or trademarks at the applicable Government / Intergovernmental Agencies (Tietze *et al.*, 2020).

[9] Dyson, which designed and built the CoVent ventilator (Jack, 2020), and Roche, which developed the Sars-CoV-2 testing kit extending its cobas 6800/8800 systems' capabilities to become Covid-19 relevant (F. Hoffmann-La Roche AG, 2020), are both examples for this new entrant group following entry strategy B. Dyson relied on its expertise with managing air flows, robotics, electronics and software (The James Dyson Foundation, 2020), as well as the expert advice from The Technology Partnership (Jack, 2020), to rapidly design and manufacture a fully functional ventilator ready for regulatory approval. Roche used its experience with developing rapid clinical diagnostics equipment to develop a PT/PCR test specifically for diagnosing Sars-CoV-2.

[10] Dyson has accumulated a considerable amount of expertise related to the efficient management of internal air flow in devices captured in both formal and informal IP (4iP Council, 2016; Justia, 2020). This complementary background IP enabled Dyson to develop a ventilator design, namely CoVent, which potentially surpasses the efficiency and performance of incumbent ventilator designs. Furthermore, Roche's knowledge base with rapid diagnostics equipment meant that it possessed complementary background IP on which basis it was able to develop a rapid Sars-CoV-2 testing solution for clinical use in the CC-Sector. With a capacity exceeding 4,000 test results in 24 hours (F. Hoffmann-La Roche AG, 2020), Roche's enhanced testing equipment for Sars-CoV-2 potentially beats the testing capabilities of Incumbent CC-Product Manufacturers and service providers.

[11] Returning to the Dyson CoVent example, the design team consciously decided to avoid traditional ventilator components in order to avoid supply chain bottlenecks that were encountered by incumbent firms (Bottomley, 2020).

Similarly to the novel informal CC-IP developed by group 4 Type 1 New Entrants in the entry strategy A scenario, the formal and informal CC-IP devised by group 3 firms in this scenario does also not qualify as either foreground or sideground IP due to the absence of any collaboration between new entrants and incumbents and the omission of any freedom-to-operate analysis by the new entrants (Granstrand and Holgersson, 2014).

As shown in Figure 5, we apply the new definition of CC-IP (paraground) here again to capture the innovation behind CC-Product enhancements introduced by group 3 Type 1 New Entrants. The pandemic CC-Sector innovation ecosystem map also captures the possibility that group 3 firms rely on tangible and intangible component inputs, such as Subassemblies from Suppliers, if dictated by the inherent complexity of the Enhanced CC-Product (Hobday, Davies and Prencipe, 2005). In this scenario, Incumbent CC-Product Manufacturers do not merely have to cope with a limited matching copy of their CC-Products in regionally limited markets as in the entry strategy A scenario but face the potent risk of a new entrant effectively becoming a potent competitor with an innovatively enhanced CC-Product, thus remaining in the CC-Sector permanently (Tietze *et al.*, 2020)

Discussed from the entry strategy A scenario, it is unlikely and of little value to Incumbent CC-Product Manufacturers to respond to group 3 Type 1 New Entrants pursuing entry strategy B during the pandemic phase due to potential reputational backlash, such as being viewed as overly exploitative during times of crisis, and because the crisis-dependent upsurge in demand for CC-Products overwhelms incumbent firms' production capacities (Tietze *et al.*, 2020). However, when the pandemic phase terminates and demand for CC-Products declines, incumbent firms may need to react to the then appearing competitive threat posed by group 3 Type 1 New Entrants in order to at least conserve their pre-pandemic market share and sales figures for CC-Products in the face of a contracting CC-Sector.

In this paper, we consider two extreme perspectives of competitive dynamics (Chen and Miller, 2015), namely the rivalrous view manifested in a defensive counter strategy on one side and the relational view represented by an open innovation strategy on the other side, to visualise how IP dynamics change in the CC-Sector innovation ecosystem after the Covid-19 pandemic terminates and group 3 Type 1 New Entrants decide to remain in the CC-Sector.

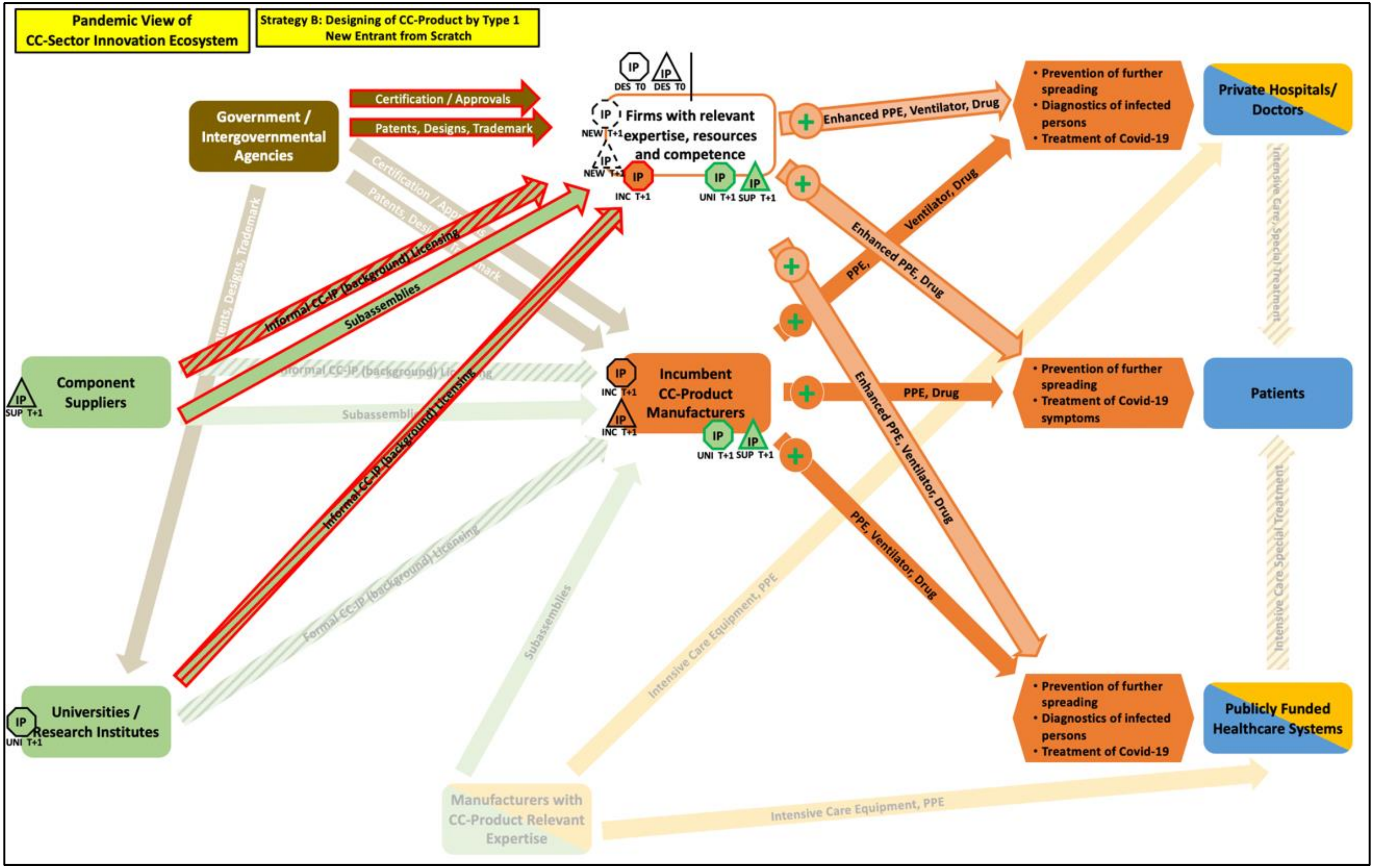

Figure 5: Visualisation of pandemic CC-Sector innovation ecosystem with group 3 Type 1 New Entrants designing CC-Products from scratch

#### 3.2.3.1 Post-pandemic response option (1) - Rivalrous Perspective: Incumbents' Defensive Counter Strategy

Incumbent CC-Product Manufacturers with a rivalrous perspective on competitive dynamics would exhibit a tendency towards short-term economically targeted tactics (Chen and Miller, 2015). As shown in our visualisation of the post-pandemic CC-Sector innovation ecosystem in Figure 6, these incumbent firms would directly attack group 3 Type 1 New Entrants for using their CC-IP (background) without authorisation by filing an infringement claim and seeking an injunction against Enhanced CC-Products as well as damage compensation, thus stimulating IP-related dynamics in the innovation ecosystem.

As stated previously, group 3 Type 1 New Entrants would typically not have conducted a freedom-to-operate analysis while designing their Enhanced CC-Products from scratch due to the urgency of meeting the positive demand shock at the start of the pandemic phase (Tietze *et al.*, 2020), which means that the likelihood of incumbents' infringement claims being successful is high. If indeed successful, group 3 Type 1 New Entrants face the prospect of having to stop selling its Enhanced CC-Products to any customer group, compensating the claiming Incumbent CC-Product Manufacturer for lost profits or at least a reasonable royalty, and losing any formal CC-IP (paraground), particularly patents due to the incumbents' CC-IP (background) effectively representing prior art (Poltorak and Lerner, 2011). Here, the legal concept of doctrine of equivalents may become particularly relevant because it stipulates that an infringement has occurred if the Enhanced CC-Product designed from scratch performs the same function, in the same way, with similar results as the incumbents' CC-Product (Poltorak and Lerner, 2011, p. 108).

Discussed from the entry strategy A scenario, Incumbent CC-Product Manufacturers need to balance the risk of high litigation costs versus the potential economic benefit of damage compensation and injunction (Poltorak and Lerner, 2011, p. 113). Contrary to Agile Manufacturers with Particular Skills in the entry strategy A scenario, group 3 Type 1 New Entrants following entry strategy B have the resources to manufacture enhanced CC-Products on a larger scale meaning that pursuing litigation may be an economically viable solution for incumbents due to damage compensation scaling with Enhanced CC-Product sales by group 3 firms (Poltorak and Lerner, 2011, p. 119).

However, there is more to consider for Incumbent CC-Product Manufacturers: access to the potentially valuable CC-IP (paraground) and complementary background IP underlying CC-Product enhancements introduced by group 3 Type 1 New Entrants. More specifically, incumbents' defensive counter strategy may prevent them from accessing background knowledge necessary to implement CC-Product enhancements developed by group 3 firms because the latter are unlikely to offer incumbent firms authorised access to its complementary background IP and remaining informal CC-IP (paraground) after being prosecuted by them. Essentially, following a defensive counter strategy has the potential to lead to short-term economic benefits for Incumbent CC-Product Manufacturers, but is unlikely to enable them to access the IP underlying CC-Product enhancements introduced by group 3 Type 1 New Entrants during the pandemic phase.

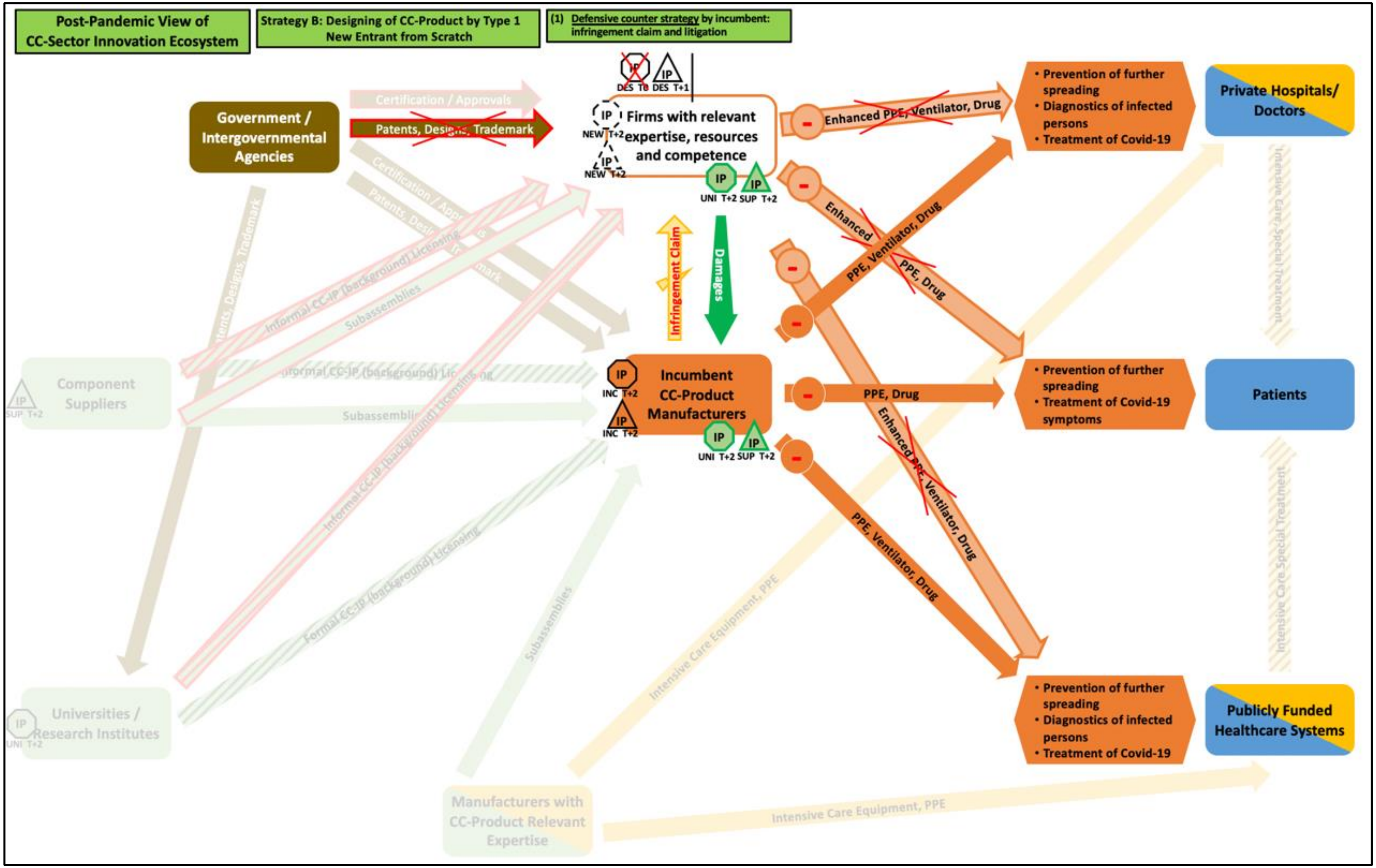

Figure 6: Visualisation of post-pandemic CC-Sector innovation ecosystem with Incumbent CC-Product Manufacturers pursuing a defensive counter strategy against group 3 firms

Group 3 Type 1 New Entrants face considerable economic risk when confronted by Incumbent CC-Product Manufacturers pursuing a defensive counter strategy. As stated above, the financial repercussions are likely to entail the costs of litigation, compensation of the incumbent for lost profits or a reasonable royalty and the potentially high development costs sunk in designing the Enhanced CC-Product from scratch due to an injunction. The group 3 Type 1 New Entrants should therefore prudently conduct a freedom-to-operate analysis as early as practicably reasonable during the crisis (Tietze *et al.*, 2020) in order to identify features of its Enhanced CC-Products and relevant formal CC-IP (paraground) potentially infringing incumbents' formal CC-IP (background)[12]. Ultimately, this would allow group 3 firms following entry strategy B to take corrective action if necessary and to design around incumbents' CC-IP (background), thereby proactively avoiding an infringement claim and litigation, as well as conserving its potential to market and sell its Enhanced CC-Product.

#### 3.2.3.2 Post-pandemic response option (2) - Relational Perspective: Incumbents' Open Innovation Strategy

Incumbent CC-Product Manufacturers with a relational perspective on competitive dynamics would focus on long-term success and appreciate that this can only be achieved if needs, preferences and interdependencies of other stakeholders in an innovation ecosystem are taken into account (Chen and Miller, 2015). In the context of the post-pandemic CC-Sector, incumbents with a relational perspective appreciate the value of macro-level technological discontinuities brought into the CC-Sector by group 3 Type 1 New Entrants and accept that access to innovations captured in CC-IP (paraground) could benefit themselves as well as the various customer groups by improving the CC-Product's value proposition. As depicted in our visualisation of the post-pandemic CC-Sector innovation ecosystem in Figure 7, incumbents with a relational perspective would be more inclined towards pursuing a collaborative, that is open innovation, approach with group 3 firms, instead of pursuing an infringement claim and compensation for lost profits or royalties.

More specifically, Incumbent CC-Product Manufacturers could proactively and restrictively (for the duration of the pandemic) license its formal CC-IP (background) to group 3 Type 1 New Entrants, thereby granting them authorised access and a limited freedom-to-operate under the terms and conditions of a license. In trade for that, group 3 Type 1 New Entrants would cross-license its formal CC-IP (paraground) to Incumbent CC-Product Manufacturers, thereby granting reciprocal authorised access under certain terms and conditions of a license. Here, Figure 7 highlights unfolding IP-related dynamics induced by the reciprocal cross-licensing between the actors, namely both the complementary background IP which group 3 firms already owned prior to the pandemic and the CC-IP (paraground) which it developed during the pandemic phase now becoming its cumulative portfolio of background CC-IP.

Typical license restrictions limiting the usage of IP are captured in change-of-technology, change-of-control, no-challenge and termination clauses (Granstrand and Holgersson, 2014), as well as scope-restriction clauses limiting the scope of usage of licensed IP to certain markets, products and customer groups. How each of these clauses are formulated and how the added value of the CC-Product enhancement is shared between the Incumbent CC-Product Manufacturers and group 3 Type 1 New Entrants is subject to IP contracting and licensing negotiations between the actors. Ahead of these negotiations, it is worthwhile for incumbent firms to remember that they can create leverage over new entrants by emphasising the alternative option of switching to a defensive counter strategy as previously described if a cross-licensing agreement under agreeable terms and conditions cannot be reached.

[12] Dyson's group IP director, Gill Smith, stated that the company takes formal IP and patents in particular seriously, and expects other companies to respect its IP portfolio in the same way it honours IP ownership of others (4iP Council, 2016). From this statement, it is reasonable to assume that Dyson has internal IP processes in place enabling it to ensure proper freedom-to-operate prior to bringing the CoVent ventilator to market.

When negotiating cross-licensing agreements, Incumbent CC-Product Manufacturers could use their bargaining power, which results from the alternative option to switch to a defensive counter strategy, to assert scope-restriction clauses that limit group 3 Type 1 New Entrants' usage of its CC-IP (background). For instance, this could limit the range of customer groups to which group 3 firms may sell Enhanced CC-Products or preclude them from independently developing any further CC-Product enhancements thereby effectively prohibiting the generation of novel CC-IP (sideground). While it would be sensible for Incumbent CC-Product Manufacturers to limit group 3 Type 1 New Entrants' access to customer groups, thereby effectively controlling their market share in the post-pandemic CC-Sector, it could potentially be beneficial not to overly restrict group 3 firms' usage of its CC-IP (background) for the development of further CC-Product enhancements in order to conserve the innovativeness of the collaboration. Instead, Incumbent CC-Product Manufacturers should use their bargaining power to assert more advantageous change-of-technology terms and conditions. For example, incumbents could insist on assign-back clauses, which force group 3 Type 1 New Entrants to transfer ownership of novel CC-IP (sideground) underlying further CC-Product enhancements during the post-pandemic phase to Incumbent CC-Product Manufacturers (Granstrand and Holgersson, 2014). Equipped with a strong negotiation mandate, Incumbent CC-Product Manufacturers can dictate favourable cross-licensing terms and conditions, whereas caution should be practiced by incumbent firms not to stifle the innovativeness of the collaboration by asserting clauses disincentivising any further CC-Product enhancements by group 3 Type 1 New Entrants.

Group 3 Type 1 New Entrants, on the other hand, may need to compromise heavily in the face of incumbents' bargaining power and to avoid the economic risk that may result from their defensive counter strategy. Considering incumbent firms' focus on scope-restriction and change-of-technology terms and conditions in cross-licensing negotiations as discussed above, group 3 firms may need to compromise between access to customer groups in the CC-Sector innovation ecosystem and ownership of novel CC-IP (sideground) underlying CC-Product enhancements that it introduces during the post-pandemic phase. If group 3 Type 1 New Entrants decide that access to a wide range of customer groups represents a more promising path to achieving their return on investment objectives, they could accept any assign-back clauses required by Incumbent CC-Product Manufacturers, thereby trading ownership of all novel CC-IP (sideground) underlying future CC-Product improvements in for access to a wider scope of customer groups. Otherwise, group 3 firms would be advised to trade access to customer groups in the CC-Sector innovation ecosystem in for replacing any assign-back clauses with grant-back clauses, retaining ownership of all novel CC-IP (sideground), but having to license it to Incumbent CC-Product Manufacturers (Granstrand and Holgersson, 2014).

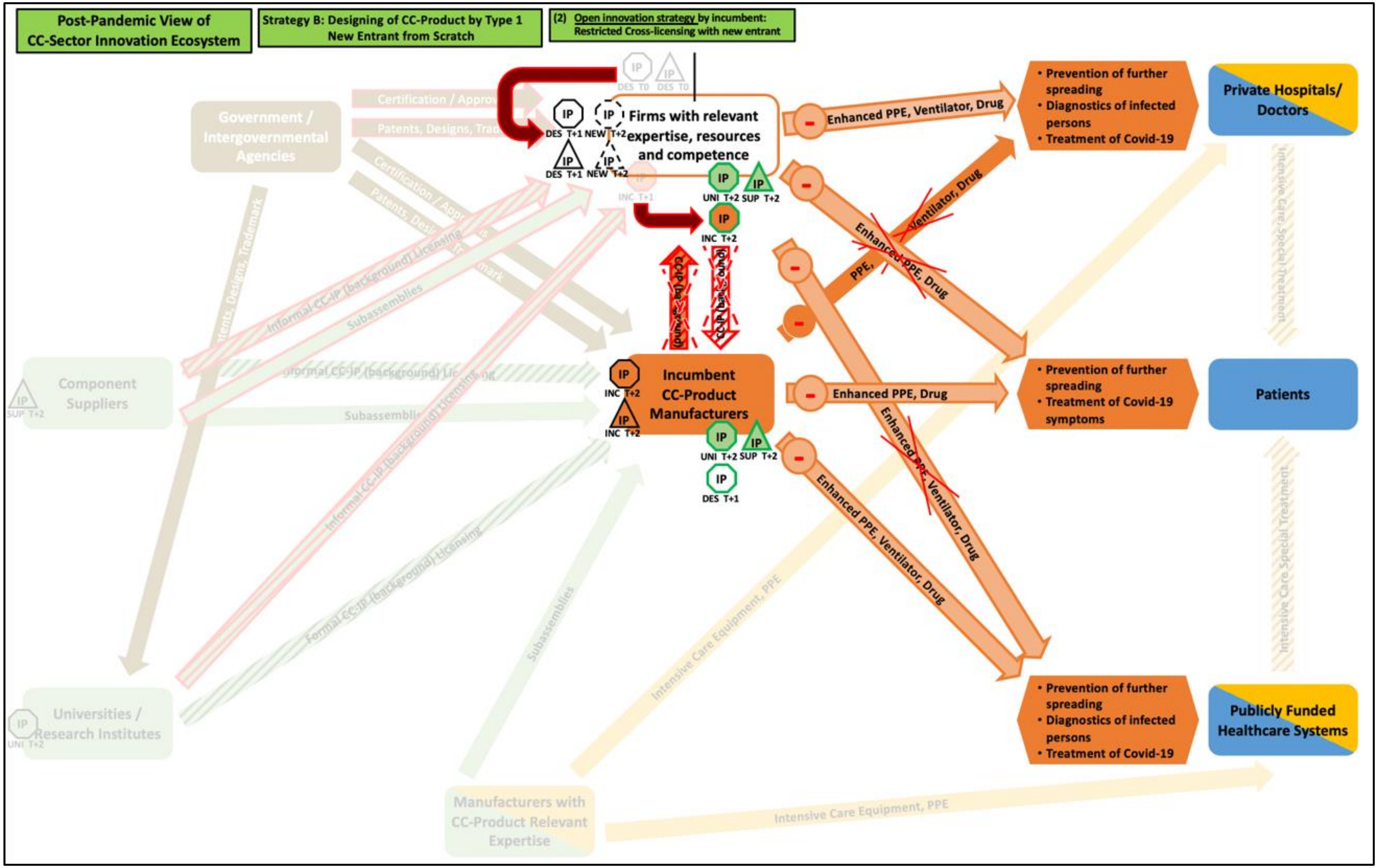

Figure 7: Visualisation of post-pandemic CC-Sector innovation ecosystem with Incumbent CC-Product Manufacturers pursuing an open innovation strategy with group 3 firms

Ultimately, group 3 Type 1 New Entrants need to decide whether they are more likely to maximise their earning potential as competitors to the Incumbent CC-Product Manufacturer by continuing to sell Enhanced CC-Products in the CC-Sector, or as a supplier of intangible goods and services, namely its cumulative portfolio of background CC-IP and novel CC-IP (sideground).

#### 3.2.4 Entry Strategy C – Teaming Up

By pursuing entry strategy C, Type 1 New Entrants repurpose their spare and complementary manufacturing resources and collaborate with Incumbent CC-Product Manufacturers that are willing to share their CC-IP (background) (Tietze *et al.*, 2020) in order to rapidly respond to the overwhelming surge in demand for CC-Products during the pandemic phase of the Covid-19 induced crisis (López-Gómez *et al.*, 2020). Type 1 New Entrants following this teaming-up approach can be characterised as either being firms with spare manufacturing capacity with related resources and capabilities, hereinafter denoted as group 1 Type 1 New Entrants, or tech giants possessing a diverse and rich set of capabilities that allows them to manufacture any product provided that access to the relevant IP is available, from now on referred to as group 2 Type 1 New Entrants (Tietze *et al.*, 2020)[13]. As exemplified by our visualisation of post-pandemic CC-Sector innovation ecosystem in Figure 8, incumbent manufacturers in this scenario typically facilitate the upscaling effort by pragmatically granting access to its CC-IP (background) to group 1 and 2 Type 1 New Entrants willing to repurpose underutilised and related manufacturing capacities due to the urgency of meeting the rapidly increasing demand for CC-Products. During the pandemic phase, group 1 and 2 Type 1 New Entrants typically experience a rapid drop in demand for their non-CC-Products[14], which they have been producing in large numbers during the pre-pandemic phase (Tietze *et al.*, 2020). Joining the CC-Sector by teaming up with incumbent firms allows group 1 and 2 firms to retain some degree of utilisation of its workforce and facilities by manufacturing CC-Products, thereby at least partially covering fixed costs and bridging the demand gap during the Covid-19 induced crisis. However, the degree of complementarity between group 1 and 2 firms' capabilities and CC-Product manufacturing requirements, as well as their skills gap relative to CC-Sector standards, must be within reasonable limits[15] in order to allow group 1 and 2 firms to recover the cost of manufacturing repurposing (López-Gómez *et al.*, 2020).

---

[13] An illustrative example for entry strategy C and group 1 and 2 Type 1 New Entrants teaming up with an Incumbent CC-Product Manufacturer is the partnership between General Motors (GM) (as Type 2 New Entrant) and Ventec Life Systems (Ventec) (as Incumbent), in which GE's former's Kokomo facility in Indiana, which is specialised in the production of precision electrical components for automobiles, was repurposed to rapidly scale manufacturing of the latter's VOCSN critical care ventilator that was jointly developed by both companies for clinical use (Brooks and Flores, 2020).

[14] Non-CC-Sectors that are affected most by the Covid-19 induced crisis and who experienced a rapid drop in demand for non-CC-Products include commercial aerospace (Bruno, 2020), automotive (BusinessWire, 2020) and luxury (Girod, 2020).

[15] GM chose its Kokomo, Indiana, facility for manufacturing the VOCSN critical care ventilator because it was idled since the start of the Covid-19 induced crisis due to the massive drop in demand for cars, thus offering the capacity, and is specialised in manufacturing electrical components for cars and offers a clean room environment, thereby providing the complementary capabilities required by the CC-Sector (Boudette and Jacobs, 2020).

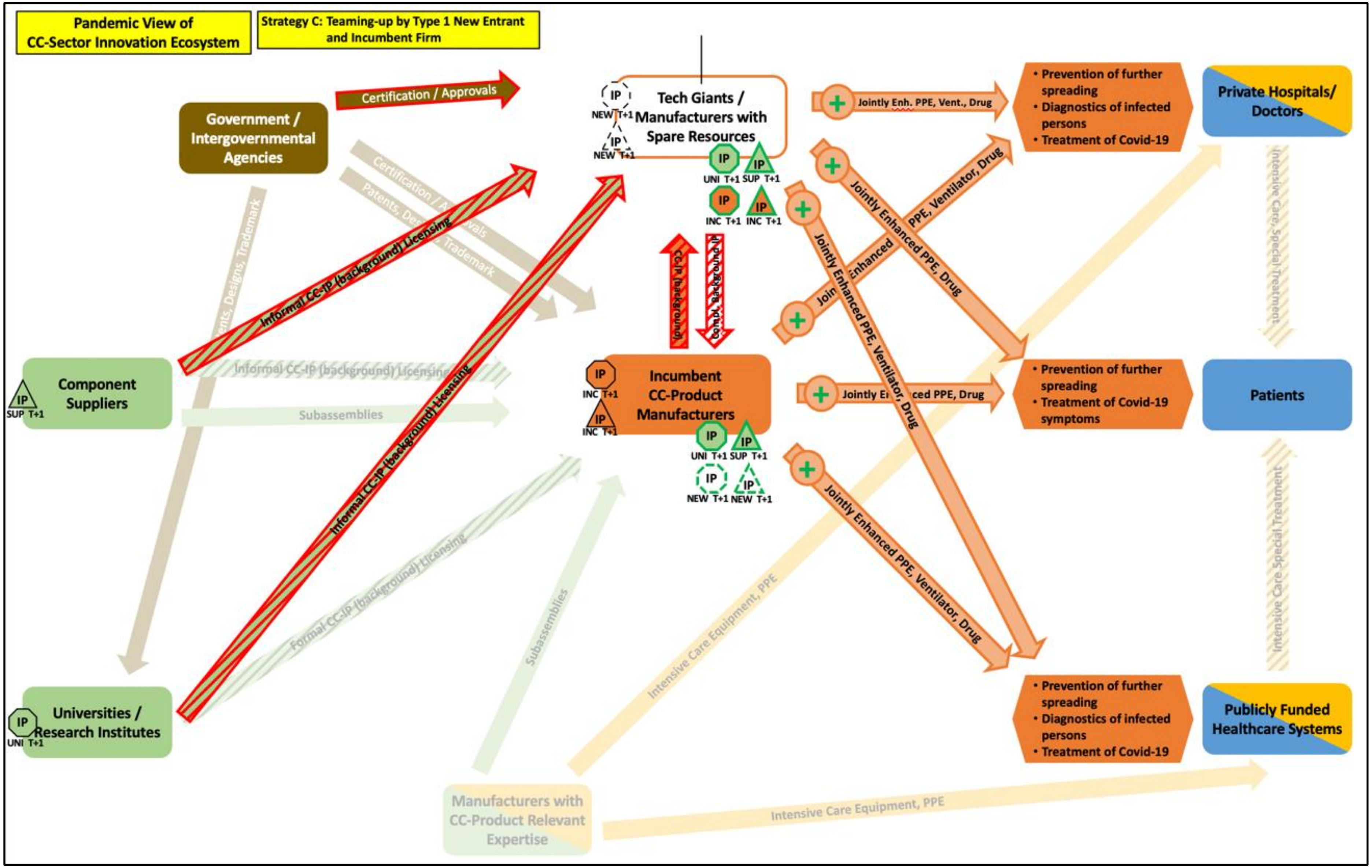

Figure 8: Visualisation of pandemic CC-Sector innovation ecosystem with group 1 (firms with spare manufacturing resources) and group 2 (tech giants) Type 1 New Entrants teaming-up with Incumbent CC-Product Manufacturers

The entry strategy C scenario represents a win-win situation for all actors involved in the collaboration because Incumbent CC-Product Manufacturers gain access to additional manufacturing capacity that helps responding to the positive demand shock for CC-Products and group 1 and 2 Type 1 New Entrants can bridge the demand gap for their non-CC-Products and utilise idle manufacturing capacities (while gaining some reputational benefits by contributing to ending the pandemic). In addition, the cross-licensing shown in Figure 8, allows group 1 and 2 Type 1 New Entrants to gain authorised access to CC-IP (background) underlying CC-Products developed and owned by incumbent firms, while also enabling group 1 and 2 firms to diffuse their complementary background knowledge into the CC-Sector[16]. Ultimately, group 1 and 2 Type New Entrants and Incumbent CC-Product Manufacturers face an IP assembly problem at the start of the collaboration (Granstrand and Holgersson, 2013), namely the challenge of integrating the necessary knowledge, technologies and skills, as well as the underlying formal and informal background IP through respective licensing agreements, in order to ensure the joint freedom-to-operate for effective scale-up of CC-Product manufacturing.

Before the collaboration effectively starts and Incumbent CC-Product Manufacturers solve the IP assembly problem together with group 1 and 2 Type 1 New Entrants, the actors need to agree on the terms and conditions of the collaboration and capture these in the aforementioned agreements, such as a reciprocal cross-licensing contracts. As already described above incumbent firms in the entry strategy C scenario provide access to their CC-IP (background) by offering some form of licenses, probably for a limited term (Tietze *et al.*, 2020), and typically expect the same from group 1 and 2 firms with respect to their complementary background IP as shown in Figure 8. While the implementation of fair and reasonable royalties rather simplifies negotiations of licensing agreements in innovation ecosystems (Granstrand, Holgersson and Opedal, 2020), defining termination and scope-restriction clauses is not as trivial due to the challenge of designing contractual mechanisms and limits ex-ante that anticipate the end of the pandemic phase and magnitude of CC-Product demand during the Covid-19 crisis, respectively. Regarding the termination clause, defining a fixed term with options to extend the collaboration upon mutual agreement or linking the term to total sales figures for CC-Products are potential solutions[17], whereas the latter may raise issues from a competition and antitrust law perspective (Nicholson Price II, Rai and Minssen, 2020). Regarding scope-restrictions, limiting usage of licensed IP to certain markets, products and customer groups could be an option as already highlighted in the entry strategy B scenario[18]. In the end, termination and scope-restriction clauses remain contentious subjects in cross-licensing negotiations because, on one hand, group 1 and 2 Type 1 New Entrants are motivated to maximise both until the cost of repurposing spare manufacturing capacities is recovered, while, on the other hand, Incumbent CC-Product Manufacturers tend to be as restrictive as possible in order to curtail potential competition in the post-pandemic phase. This certainly calls for some involvement of experienced negotiators, who can act quick, agile to strike some potentially unconventional deal.

[16] In the example of the scaling up manufacturing of the VOCSN critical care ventilator, GM brings its global supplier base, vast logistics network, manufacturing, engineering, purchasing and legal prowess, all of which hold GM's complementary formal and informal background IP, into the collaboration in order to meet the challenging goal to source the more than 700 individual parts according to Ventec's specifications, which represents its CC-IP (background), necessary to assemble each of the planned two hundred thousand VOCSN ventilators (Brooks and Flores, 2020).

[17] This was raised by an NGT expert based upon several years of experience with negotiating licensing agreements with large OEMs in the commercial aerospace sector.

[18] Ventec, for instance, took an extreme stance towards scope-restriction and insisted on remaining the primary point of sale and distribution for VOCSN critical care ventilators manufactured through the collaboration with GM, despite its small size compared to its partner (Boudette and Jacobs, 2020).

As the term of the cross-licensing agreements and hence the collaboration is likely to be time limited (Tietze *et al.*, 2020), Incumbent CC-Product Manufacturers and group 1 and 2 Type 1 New Entrants will analogously face an IP disassembly problem (Granstrand and Holgersson, 2013), namely the challenge of disentangling jointly developed knowledge, technologies and skills, as well as the underlying formal and informal IP, through appropriate contingency clauses towards the end of the pandemic phase. More specifically, the transfer and recombination of CC-IP (background) owned by Incumbent CC-Product Manufacturers and complementary background IP held by group 1 and 2 Type 1 New Entrants potentially leads to macro-level technological discontinuities and, ultimately, to really new innovations (Garcia and Calantone, 2002) and CC-Product enhancements, such as reduced manufacturing costs (Tietze *et al.*, 2020), reduced lead-time or improved CC-Product quality enabled by new standardized best practices (Nicholson Price II, Rai and Minssen, 2020). These CC-Product enhancements could be captured either as CC-IP (foreground) if it is developed collaboratively by incumbents and group 1 and 2 firms, or as CC-IP (sideground) if it is developed by either actor individually outside of the collaboration (Granstrand and Holgersson, 2014). Discussed from the entry strategy B scenario, change-of-technology clauses represent an effective contractual mechanism for solving the IP disassembly problem (Granstrand and Holgersson, 2013) by regulating the allocation of ownership and usage of both jointly developed CC-IP (foreground) and independently developed CC-IP (sideground) after cross-licensing agreements are terminated at the end of the pandemic phase. In the following two sub-sections, two choices of change-of-technology clauses that were already discussed in the entry strategy B post-pandemic context are reconsidered in order to identify their respective effects on the pandemic CC-Sector innovation ecosystem in which group 1 and 2 Type 1 New Entrants team-up with Incumbent CC-Product Manufacturers . These options are (i) grant-back and (ii) assign-back clauses.

#### 3.2.4.1 Pandemic Change-of-Technology: Grant-back Clause

If Incumbent CC-Product Manufacturers and group 1 and 2 Type 1 New Entrants agree on a grant-back clause, ownership of newly developed CC-IP that is based on the CC-IP (background) of incumbents and/or complementary background IP of group 1 and 2 firms would remain with the innovator, but under the obligation of having to license it back to the collaboration partner (Granstrand and Holgersson, 2014). As shown in our visualisation of the pandemic CC-Sector innovation ecosystem governed by a grant-back clause in Figure 9, any CC-IP (foreground) that is jointly developed by Incumbent CC-Product Manufacturers and group 1 and 2 Type 1 New Entrants on the basis of the former’s CC-IP (background) and the latter’s complementary background IP is jointly owned. Furthermore shown in Figure 9, any CC-IP (sideground) that is independently developed by group 1 and 2 Type 1 New Entrants on the basis of both their complementary background IP and incumbents’ CC-IP (background) remains in the ownership of group 1 and 2 firms under the explicit contractual condition that a license to such CC-IP (sideground) is granted to incumbent firms under the general terms and conditions of the cross-licensing agreement.

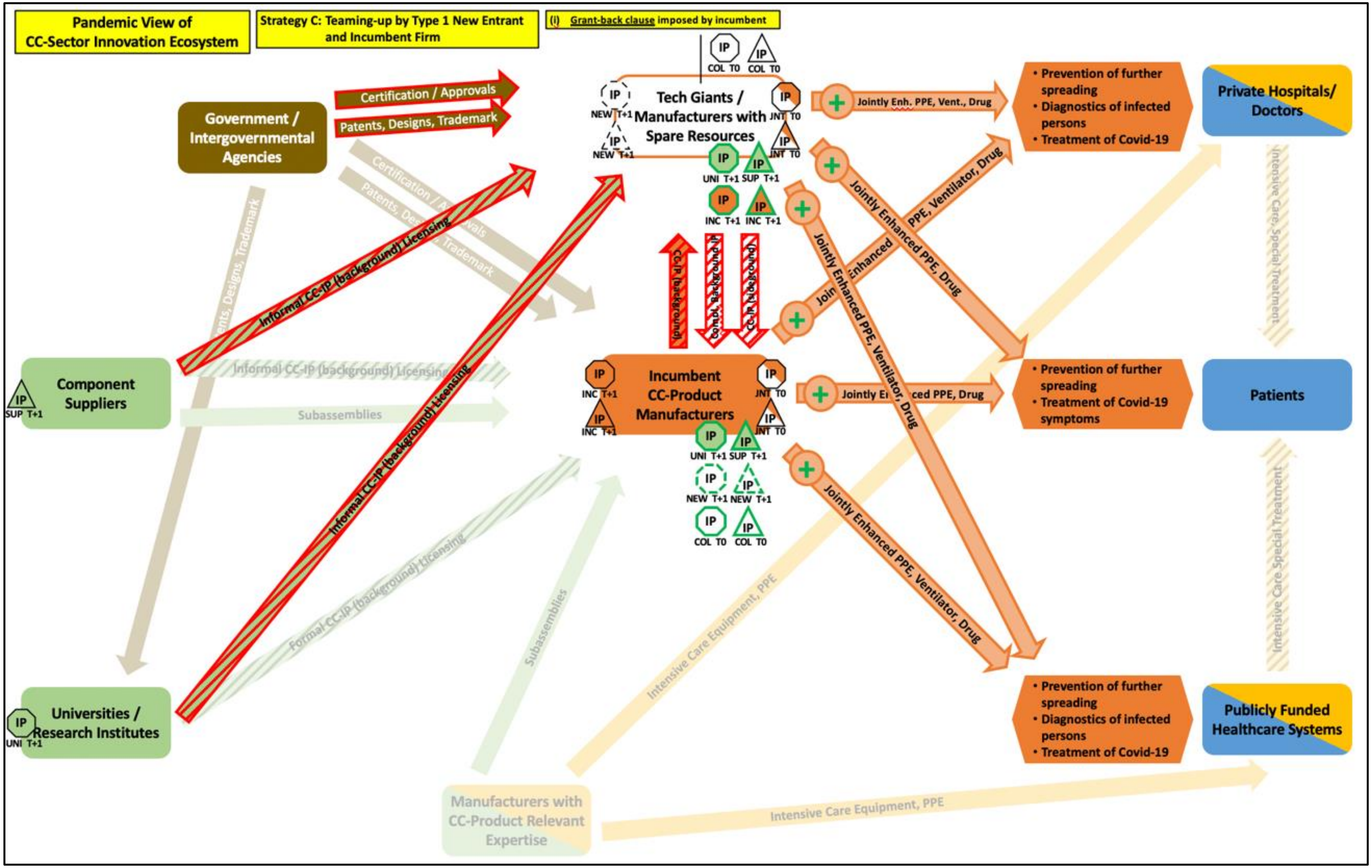

Figure 9: Visualisation of pandemic CC-Sector innovation ecosystem with group 1 and 2 Type 1 New Entrants and Incumbent CC-Product Manufacturers agreeing on a grant-back clause

Agreeing on a grant back clause ultimately allows the collaboration partners to hold on to ownership of the IP underlying their respective endogenously or collaboratively developed innovation, as the case may be, while at the same time conserving the reciprocal authorised access to background and novel CC-IP, hence the freedom-to-operate, that is necessary for both collaboration partners to supply jointly enhanced CC-Products to all customer groups.

#### 3.2.4.2 Pandemic Change-of-Technology: Assign-back Clause

If Incumbent CC-Product Manufacturers and group 1 and 2 Type 1 New Entrants agree on an assign-back clause, ownership of newly developed CC-IP is transferred back from the innovator to the collaboration partner owning the respective CC-IP (background) or complementary background IP on whose basis the newly developed CC-IP is based (Granstrand and Holgersson, 2014). Our visualisation of the pandemic CC-Sector innovation ecosystem in Figure 10 shows the IP-related dynamics induced by the assign-back clause, namely full ownership of CC-IP (sideground) and joint ownership of CC-IP (foreground) is fully transferred by group 1 and 2 Type 1 New Entrants to Incumbent CC-Product Manufacturer, while the former retains the right to license back the newly developed CC-IP under the terms of the cross-licensing agreement. While newly developed CC-IP that is based on both CC-IP (background) and complementary background IP is generally eligible to shared ownership by both incumbent and group 1 and 2 firms, Incumbent CC-Product Manufacturers might want to claim exclusive ownership of any newly developed CC-IP by new entrants because they consider it to be of core importance to its activities in the CC-Sector and consider it non-core to group 1 and 2 Type 1 New Entrants in the pandemic phase. The assign-back clause thus effectively leads to an automatic transfer of ownership of both CC-IP (foreground), which is jointly developed by incumbent and group 1 and 2 firms, and CC-IP (sideground), which is individually developed by group 1 and 2 firms, to the Incumbent CC-Product Manufacturers regardless of whether it was developed on the basis of CC-IP (background) alone or both CC-IP (background) and complementary background IP. Figure 10 below also shows that despite having to transfer ownership of all newly developed CC-IP back to the incumbent firm, the licensing back of CC-IP (foreground) and CC-IP (sideground) by incumbent firms conserves group 1 and 2 firms' full freedom-to-operate for the continued manufacturing and distribution of enhanced CC-Products to all customer groups during the pandemic phase.

As long as the cross-licensing agreement is effective during the pandemic phase, it makes little difference to either collaboration partner whether ownership of newly developed IP is governed by a grant-back or an assign-back clause because back licensing terms and conditions ensure that both group 1 and 2 Type 1 New Entrants and Incumbent CC-Product Manufacturers have freedom-to-operate, thus to manufacture and sell the jointly developed CC-Products to all customer groups as discussed above and shown in Figure 9 and Figure 10. Only as the pandemic terminates and demand for jointly developed CC-Products declines towards pre-pandemic levels, the termination clause agreed upon by the collaboration partners takes effect and ends the cross-licensing agreement. Group 1 and 2 and incumbent firms then need to decide whether to continue the collaboration and agree on post-pandemic (cross-) licensing terms and conditions or whether to end the collaboration. At this point the consequences of the choice between the two change-of-technology options become important.

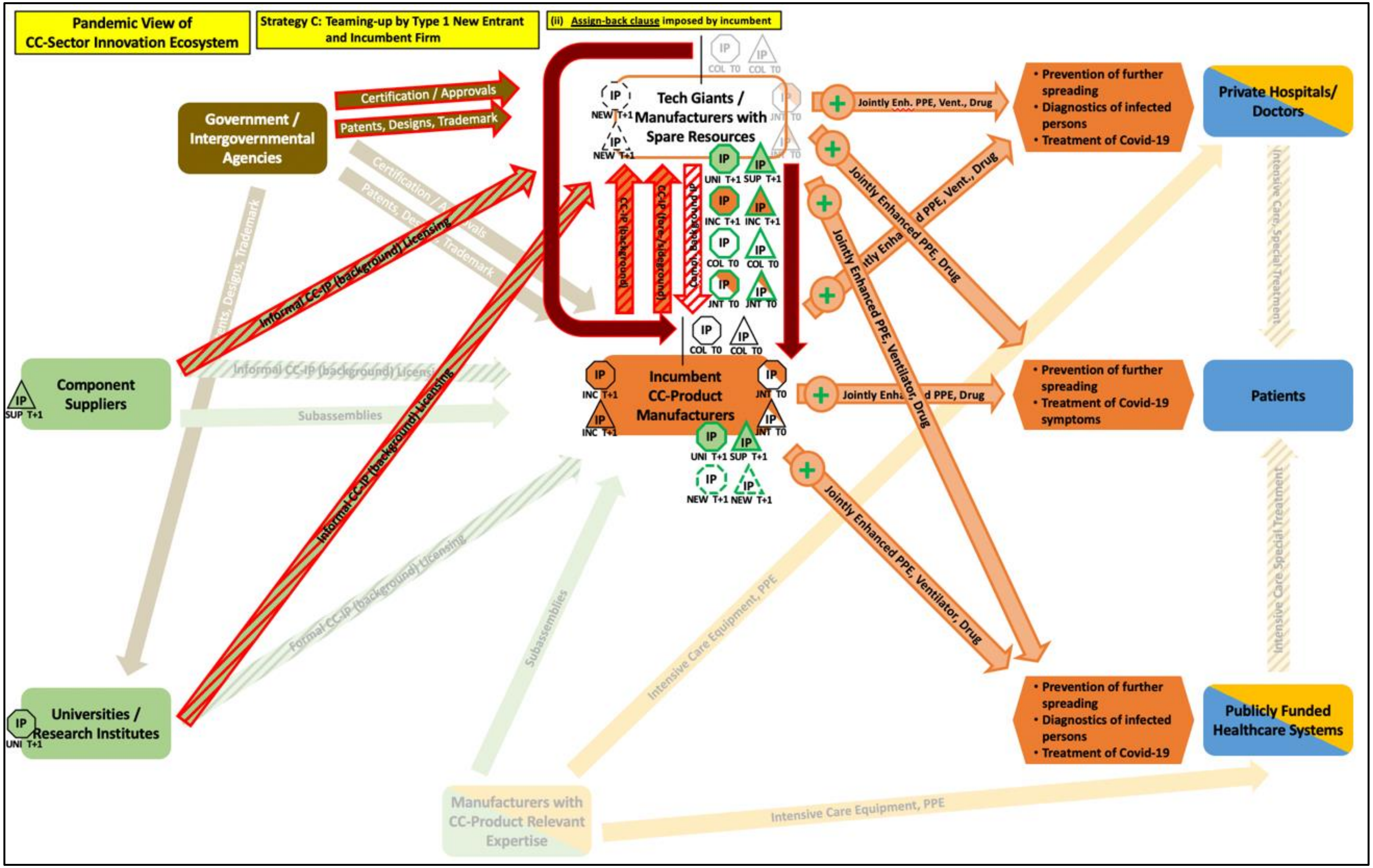

Figure 10: Visualisation of pandemic CC-Sector innovation ecosystem with group 1 and 2 Type 1 New Entrants and Incumbent CC-Product Manufacturers agreeing on an assign-back clause

In the following, we consider one post-pandemic CC-Sector innovation ecosystem in which the collaboration is terminated and another in which the collaboration is renewed in order to delineate the consequences of the change-of-technology option chosen at the start of the pandemic phase for each collaboration partner.

#### 3.2.4.3 Post-pandemic response option (3) – Rivalrous Perspective: Incumbents ending collaboration

As the pandemic phase ends, demand for CC-Products is likely to decrease to pre-pandemic levels, but contractual mechanisms defined in the termination clause of the cross-licensing agreement may stipulate a continuation beyond the end of the Covid-19 induced crisis. In order to defend their pre-pandemic market share and level of sales in this contracting CC-Sector, Incumbent CC-Product Manufacturers may choose to adopt a rivalrous perspective on competitive dynamics (Chen and Miller, 2015) similarly to the post-pandemic response option (1) in the entry strategy B scenario. This means that incumbent firms stop viewing group 1 and 2 Type 1 New Entrants as collaboration partners, but as competition instead that needs to be effectively forced out of the CC-Sector in the post-pandemic phase[19]. The obvious choice for incumbent firms therefore is to end the collaboration by expediting the termination clause, which is herein referred to the post-pandemic response option (3). From Figure 11, we can see that the end of the collaboration and hence the termination of the cross-licensing terms and conditions at the start of post-pandemic phase mean that established links for authorised IP access between former collaboration partners are undone. More specifically, as depicted in Figure 11, if IP ownership and usage during the pandemic phase was governed by a grant-back clause, the end of the licensing arrangement leads to group 1 and 2 Type 1 New Entrants losing authorised access to incumbent firms' CC-IP (background) and Incumbent CC-Product Manufacturers having to forfeit authorised access to group 1 and 2 firms' complementary background IP and CC-IP (sideground). Furthermore, co-ownership of jointly developed CC-IP (foreground) is conserved. Ultimately, both group 1 and 2 Type 1 New Entrants and Incumbent CC-Product Manufacturers lose their respective previously held freedom-to-operate with respect to manufacturing jointly Enhanced CC-Products. While incumbent firms potentially need to return to manufacturing and selling the original, pre-pandemic version of the CC-Product but potentially with selected enhanced features, depending on which features of the jointly developed Enhanced CC-Product are embedded in jointly owned CC-IP (foreground), group 1 and 2 Type 1 New Entrants have to stop manufacturing and selling jointly enhanced CC-Products altogether.

If IP ownership and usage during the pandemic phase was governed by an assign-back clause, our respective visualisation of the post-pandemic CC-Sector innovation ecosystem in Figure 12 shows that the end of the cross-licensing arrangement leads to group 1 and 2 Type 1 New Entrants basically losing authorised access to any CC-IP namely CC-IP (background) owned by incumbent firms, jointly developed CC-IP (foreground), and even endogenously developed CC-IP (sideground).

[19] Group 1 and 2 Type 1 New Entrants' interests in the post-pandemic phase may be aligned because their origin non-CC-Sector recovers and demand for their non-CC-Products, which they have been producing in large numbers prior to the Covid-19 induced crisis, increases. In addition, group 1 and 2 firms may have succeeded in recovering their investment in repurposing their complementary manufacturing capacities during the pandemic phase.

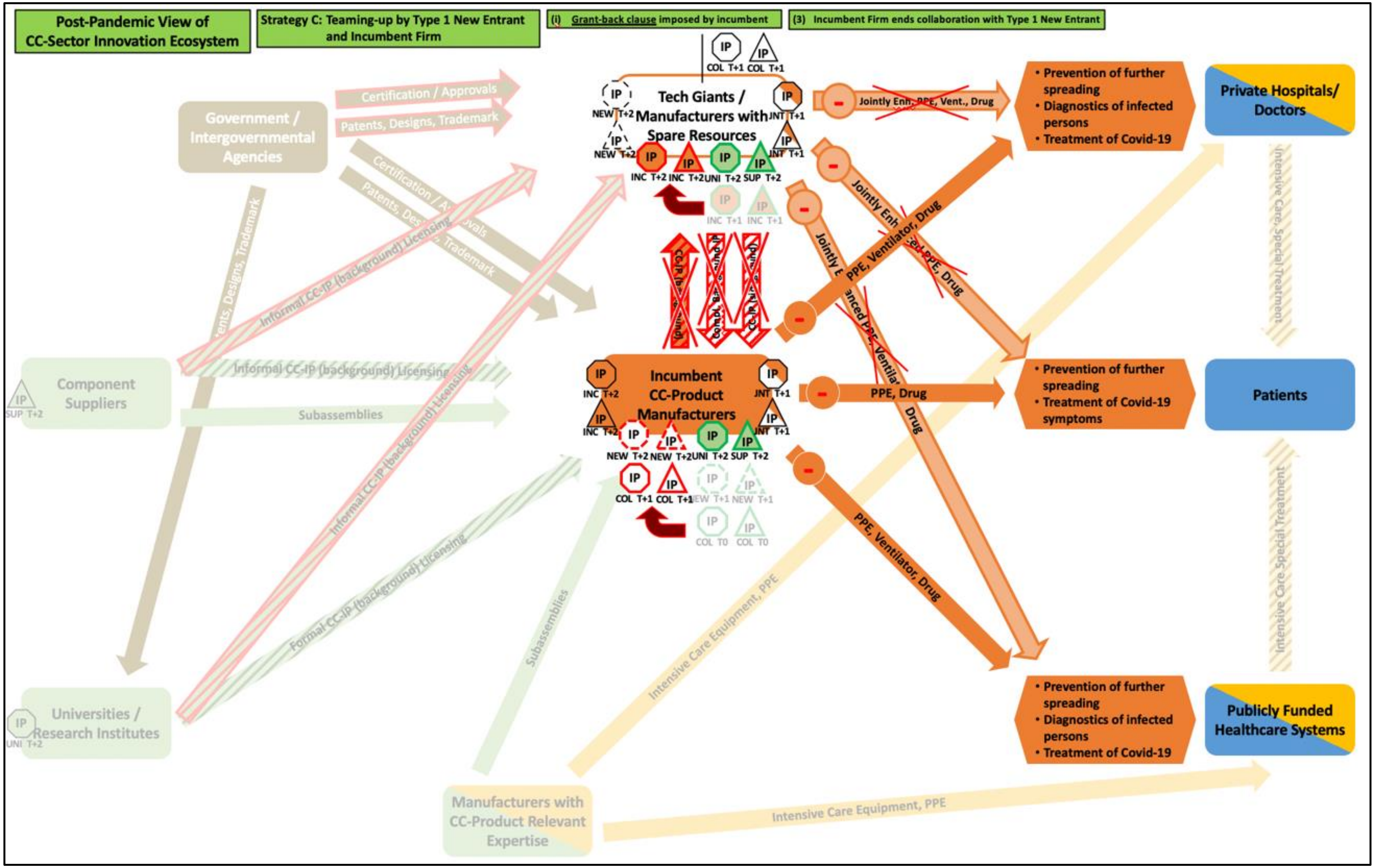

Figure 11: Visualisation of post-pandemic CC-Sector innovation ecosystem with incumbent firms ending cross-licensing agreements governed by a grant-back change-of-technology clause

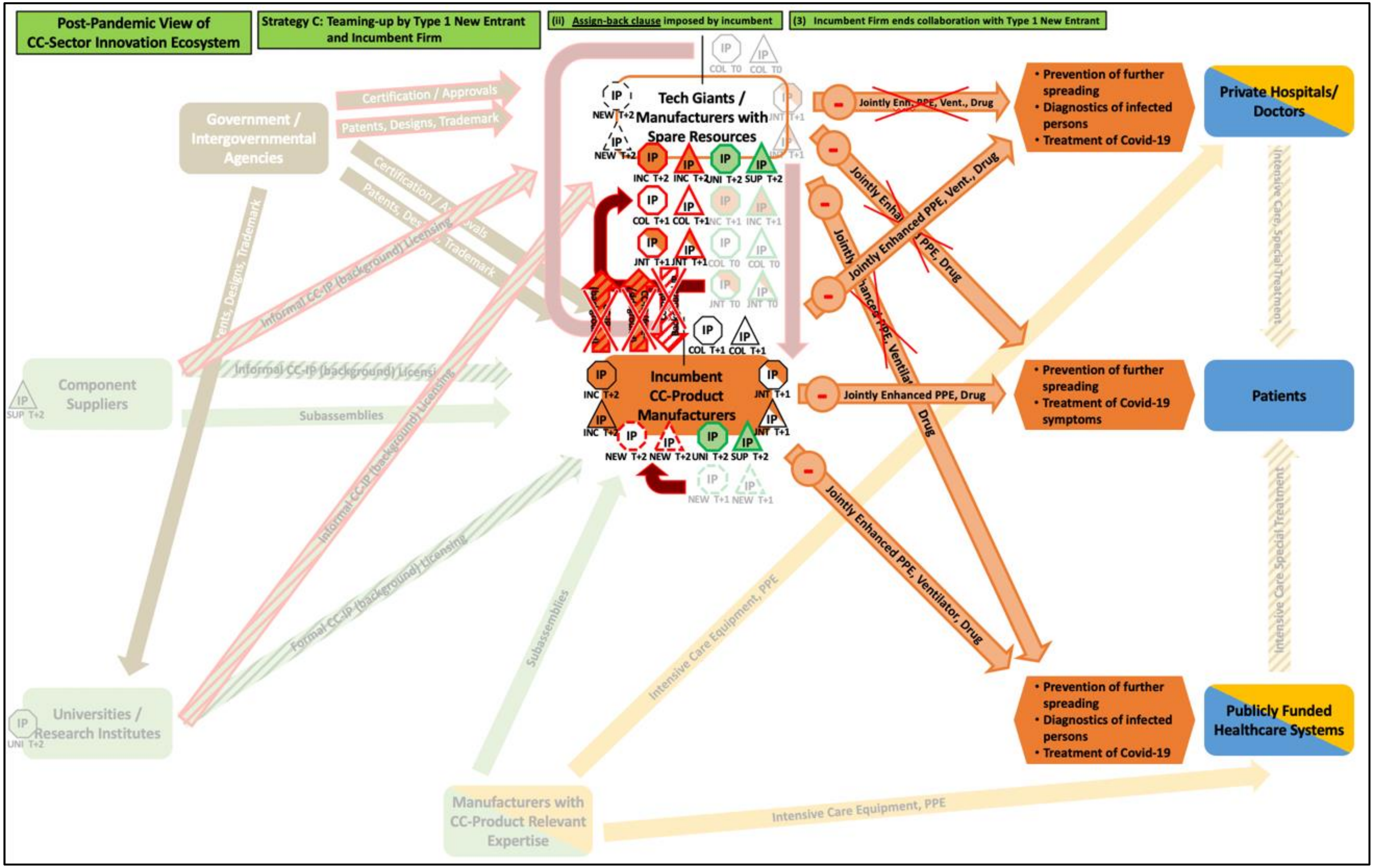

Figure 12: Visualisation of post-pandemic CC-Sector innovation ecosystem with incumbent firms ending cross-licensing agreements governed by an assign-back change-of-technology clause

Incumbent CC-Product Manufacturers, on the other hand, only forfeit authorised access to complementary background IP owned by group 1 and 2 firms. Ultimately, group 1 and 2 Type 1 New Entrants again lose their freedom-to-operate necessary to continue manufacturing jointly enhanced CC-Products, while incumbent firms would potentially be able to continue manufacturing and selling CC-Products with enhanced features resulting from joint developments during the collaboration depending on which features are captured by back-assigned CC-IP (sideground) and CC-IP (foreground).

Asserting an assign-back clause in the cross-licensing agreement during the pandemic phase might appear to be strategically advantageous from the perspective of the Incumbent CC-Product Manufacturer because it maximises the potential freedom-to-operate and allows the continuation of exclusive enhanced CC-Product manufacturing by the incumbent firm in the post-pandemic phase[20]. Furthermore, an assign-back clause avoids the co-ownership of jointly developed CC-IP (foreground) resulting from the grant-back clause as exemplified by the respective CC-Sector innovation ecosystem in
Figure 12, thus mitigating risks associated with limited control over collaboration partners' usage of such novel CC-IP (Granstrand and Holgersson, 2014). For example, there is a viable risk that a group 1 and 2 Type 1 New Entrant may license jointly owned CC-IP (foreground) to other incumbent firms in the CC-Sector in an effort to continue monetising its non-core IP portfolio, which cannot be in the interest of its former collaboration partner. From the perspective of group 1 and 2 Type 1 New entrants, agreeing on a grant-back clause during the pandemic phase appears more beneficial as it allows group 1 and 2 firms to appropriate further value from its collaboration with Incumbent CC-Product manufacturers during the pandemic phase. More specifically, as exclusively owned CC-IP (sideground) and jointly owned CC-IP (foreground) would typically capture technologies of non-core importance to group 1 and 2 firms' activities in the non-CC-Sector, continued monetisation through sub-licensing to other incumbent firms may be a viable option. Alternatively, transfer of the novel CC-IP into the non-CC-Sector could be conducted if the technology is deemed complementary to group 1 and 2 firms' core activities.

#### 3.2.4.4 Post-pandemic response option (4) – Relational Perspective: Continuation of Collaboration

As an alternative to the response option 3, incumbent CC-Product Manufacturers can adopt a relational approach to managing the post-pandemic phase in the CC-Sector and view group 1 and 2 Type 1 New Entrants as a sustainable source of further innovation resulting from the recombination of complementary background IP, CC-IP (background) and novel CC-IP developed during the pandemic phase. Similarly to the post-pandemic response option (2) in the context of the entry strategy B scenario, this perspective leads incumbent firms to continuing or renewing the collaboration with group 1 and 2 firms in the post-pandemic phase, which is hereinafter denoted as the post-pandemic response option (4)[21]. The licensing

[20] In the context of the collaboration example of GM and Ventec quoted earlier, an assign-back clause could potentially enable the latter to continuously benefit from the collaboration, for instance, by adopting newly devised VOCSN production and assembly processes implemented by GM in its repurposed Kokoma facility.

[21] While the actors' motivation to collaborate during the pandemic phase was primarily driven by the urgent call to respond to a crisis situation with a strict focus on covering costs as shown in the GM and Ventec example

terms governing authorised usage of background and novel CC-IP during the pandemic phase are, however, likely to be succeeded by different agreements, such as containing elevated royalty fees in exchange for authorised CC-IP usage in the post-pandemic phase (Tietze *et al.*, 2020). Furthermore, Incumbent CC-Product Manufacturers and group 1 and 2 Type 1 New Entrants will need to engage into an IP reassembly process, which is the recurring cycle of sequentially solving the IP disassembly and assembly problem between two temporally distinct collaboration projects (Granstrand and Holgersson, 2014), when transitioning from the ad hoc developed collaboration during the pandemic to the post-pandemic collaboration, during which the objectives of collaboration partners are likely to be different.

Our visualisation of the post-pandemic CC-Sector innovation ecosystem in Figure 13 shows the IP-related dynamics stimulated by a grant-back change-of-technology clause during the IP reassembly process at the start of the post-pandemic phase. Both collaboration partners start accumulating portfolios of background CC-IP. More specifically, group 1 and 2 Type 1 New Entrants integrate CC-IP (foreground) and CC-IP (sideground) with its complementary background IP to form its cumulative portfolio of background CC-IP, which becomes subject to royalty-based licensing terms and conditions to Incumbent CC-Product Manufacturers. Conversely, Incumbent CC-Product Manufacturers consider the jointly owned CC-IP (foreground) to become part of its cumulative portfolio of background CC-IP, to which it offers authorised access under royalty-based licensing terms and conditions to group 1 and 2 firms. The post-pandemic CC-Sector innovation ecosystem map in Figure 13 also highlights the restricted freedom-to-operate that may result from scope-restriction clauses, on which the continuing collaboration partners may agree.

If collaboration partners agreed on an assign-back clause during the pandemic phase, the IP-related dynamics during the IP reassembly process at the start of the post-pandemic phase would differ considerably as shown in our depiction of the respective post-pandemic CC-Sector innovation ecosystem in Figure 14. More specifically, group 1 and 2 firms' cumulative portfolio of background CC-IP would only consist of its complementary background IP from before the pandemic. Incumbent CC-Product Manufacturers, on the other hand, would amass a background CC-IP portfolio consisting of its endogenously developed CC-IP (background), jointly developed CC-IP (foreground) and exogenously developed CC-IP (sideground). From observing Figure 14, it becomes evident that an assign-back clause agreed in the pandemic phase effectively leads to unevenly distributed background CC-IP portfolios in the post-pandemic CC-Sector innovation ecosystem. But nevertheless, a royalty-based cross-licensing arrangement would conserve sufficient freedom-to-operate allowing both collaboration partners to continue to manufacture and sell jointly developed CC-Products.

From the perspective of Incumbent CC-Product Manufacturers, asserting an assign-back clause in the cross-licensing agreement during the pandemic phase is again more beneficial than a grant-back clause primarily because it enables incumbent firms to quickly accumulate a sizeable and exclusively owned background CC-IP portfolio[22].

---

(Boudette and Jacobs, 2020), competitive considerations with a focus on longer term value creation and win-win exchanges (Chen and Miller, 2015) become dominant in a post-pandemic setting.

[22] Furthermore, the cost and burden of research and development that formed the basis of jointly developed CC-Product enhancements was shared with group 1 and 2 Type 1 New Entrants during the pandemic phase.

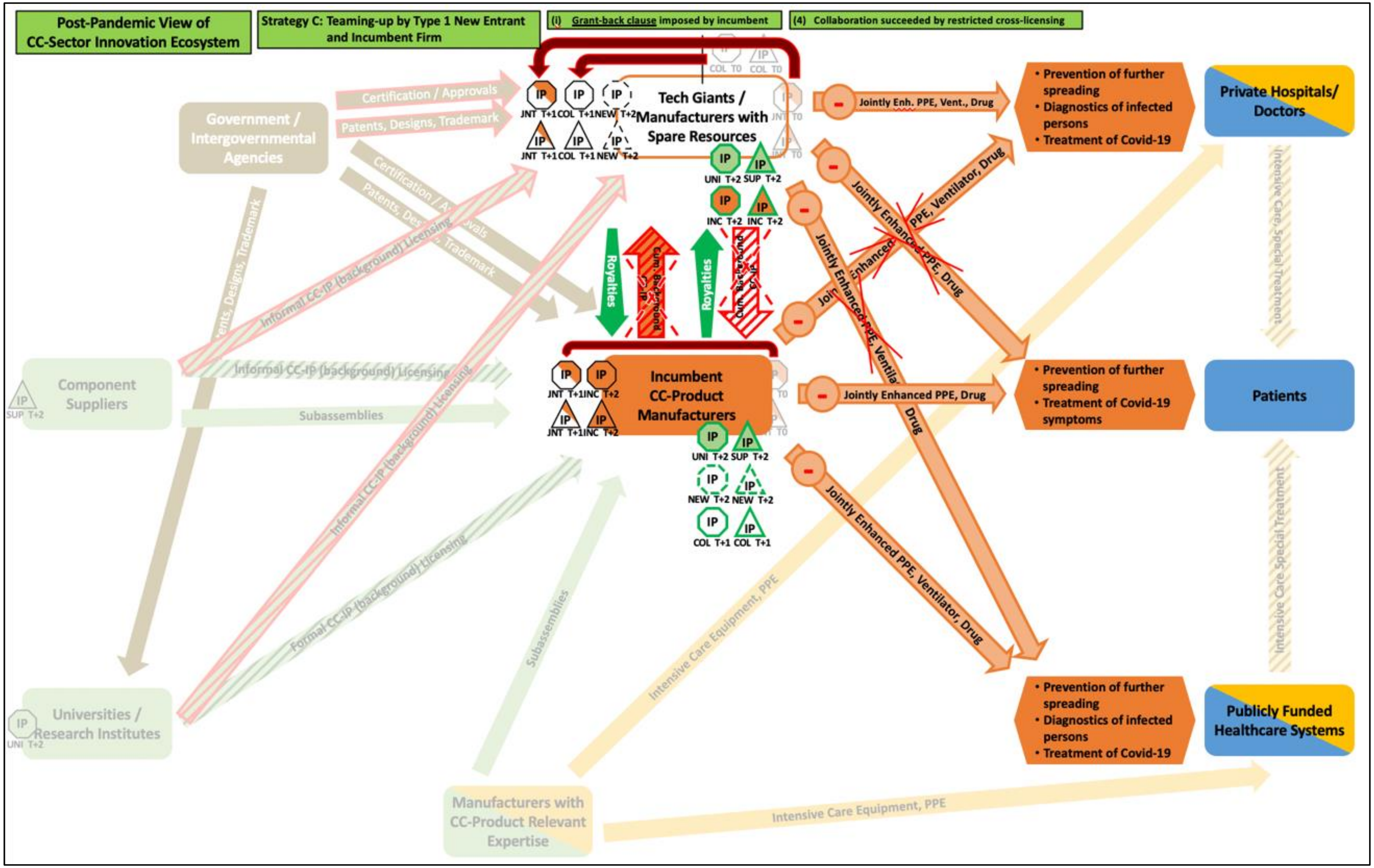

Figure 13: Visualisation of post-pandemic CC-Sector innovation ecosystem with incumbent and group 1 and 2 firms continuing cross-licensing agreements on the basis of a grant-back change-of-technology clause during the pandemic phase

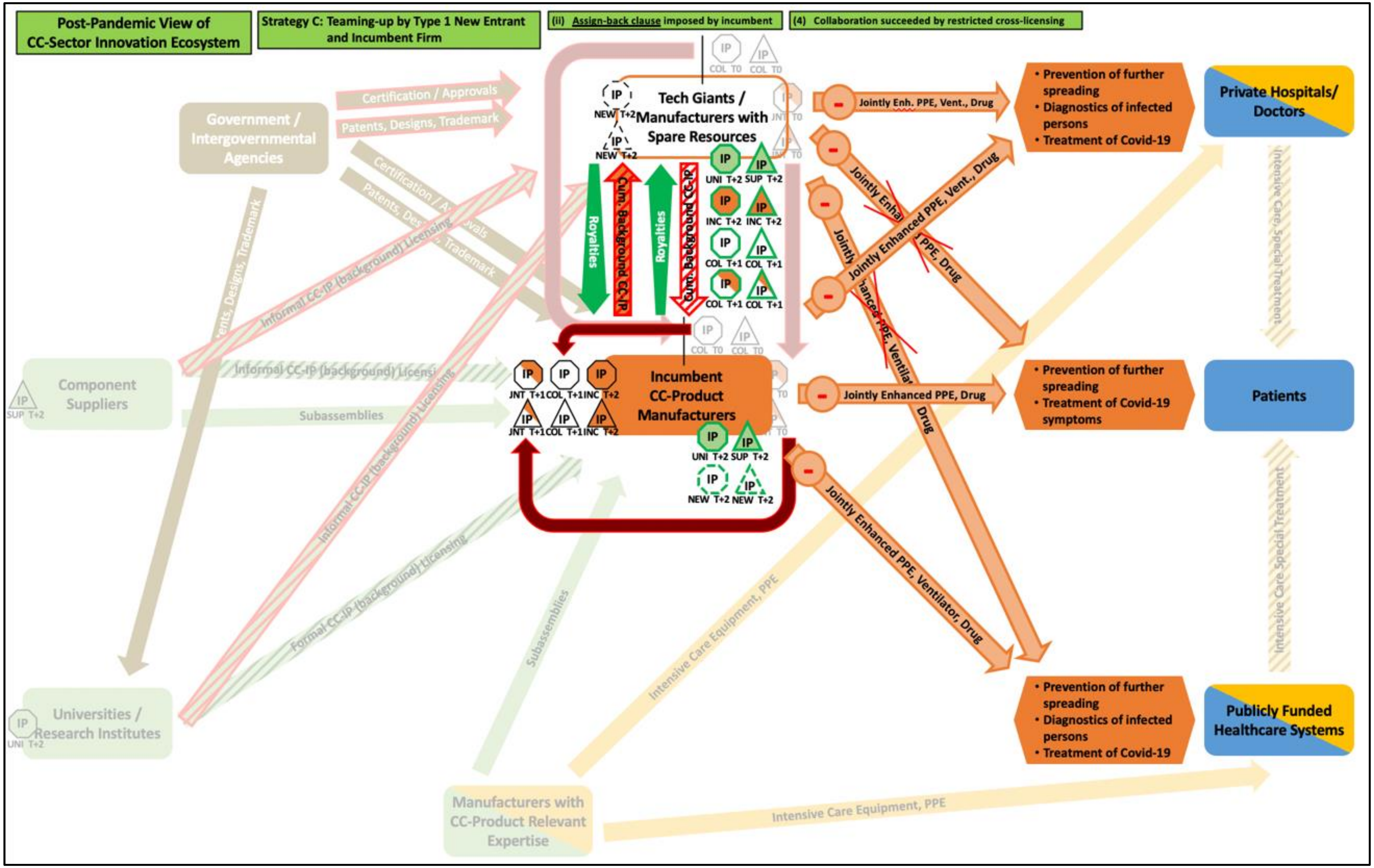

Figure 14: Visualisation of post-pandemic CC-Sector innovation ecosystem with incumbent and group 1 and 2 firms continuing cross-licensing agreements on the basis of a grant-back change-of-technology clause during the pandemic phase

The larger background CC-IP portfolio, in turn, allows incumbent firms to build bargaining power in subsequent cross-licensing negotiations with group 1 and 2 firms, particularly when asserting favourable scope-restriction clauses and royalty fees for the post-pandemic phase. As already discussed in the context of the post-pandemic response option (2) of the entry strategy B scenario, however, building and using this bargaining power can become a double-edged sword because it may irritate group 1 and 2 Type 1 New Entrants and stimulate them to disengage from negotiations of a post-pandemic licensing agreement, thus voluntarily exiting the CC-Sector innovation ecosystem. Ultimately, if incumbent firms intend to continue the collaboration with group 1 and 2 firms in the post-pandemic phase, they may need to incentivise their former and hopefully continuing collaboration partner to remain in the CC-Sector by creating a win-win situation either through less restrictive scope-restriction clauses or preferential royalty fees for authorised access to its background CC-IP portfolio.

Thus, maintaining a grant-back clause in the cross-licensing agreement during the pandemic phase appears the more advantageous strategic choice for group 1 and 2 Type 1 New Entrants because it enables them to hold on to a considerable background CC-IP portfolio. This, in turn, may even be necessary for group 1 and 2 firms in order to recoup the investment in repurposing its complementary manufacturing capacities at the start of the pandemic phase by continued monetisation of its background CC-IP portfolio in the post-pandemic phase. More importantly, however, the grant-back clause and resulting background CC-IP portfolio may create a more level playing field for the cross-licensing negotiations with Incumbent CC-Product Manufacturers. On the other hand, group 1 and 2 firms' motivation to monetise its background CC-IP portfolio combined with the aforementioned ambiguity around jointly owned CC-IP (foreground) resulting from the grant-back clause during the pandemic phase, may attract the scrutiny of incumbent firms, particularly if group 1 and 2 firms offer access to jointly developed CC-IP (foreground) to incumbents' competitors in the post-pandemic CC-Sector innovation ecosystem[23].

## 4.0 Conclusions, Limitations and Future Research

The goal of this study is to help further the understanding of IP related dynamics, risks and uncertainties in the CC-Sector as the Covid-19 induced crisis unfolds and new entrants are called into or voluntarily join incumbents in the CC-sector to help out in the effort to mitigate the crisis situation. We accomplished this by taking an innovation ecosystem perspective and adapting, as well as extending a previously developed standardised visual approach, namely the SVEL proposed by Moerchel et al. (2021), to map the CC-Sector for each of Tietze et al.'s (2020) three proposed entry strategies, namely (A) insouciantly adopting, (B) designing from scratch and (C) teaming up.

The results indicate that the threat of Type 1 New Entrants permanently establishing themselves as competing actors in the CC-Sector is particularly high in the entry strategy (B) scenario and exists to some extent also in entry strategy (C). In any case, following a defensive counter strategy by filing an infringement claim against Type 1 New Entrants, thus enforcing an injunction and seeking compensation, exposes Incumbent CC-Product Manufacturers to the risk of high litigation costs not being recouped or even failing altogether. Furthermore, this defensive approach typically implicates that incumbent firms' access to potentially valuable complementary background IP and newly developed CC-IP owned by Type 1 New Entrants is denied permanently, thus blocking the way to CC-Product innovations. On the other hand, when pursuing a collaborative approach, incumbent firms might consider to assert assign-back clauses in cross-licensing agreements in order to quickly accumulate a sizeable portfolio of CC-IP (background) while sharing the cost of its development, to avoid risks of shared CC-IP ownership, and to build up leverage for the next round of cross-licensing negotiations.

[23] Granstrand and Holgersson (2014) document a case, in which a collaboration partner sued its counterpart for breaching post-contractual loyalty, infringing its patents and misappropriating its trade secrets after being denied to search for an alternative (3rd party) collaboration partner by showing jointly developed facilities.

The study also shows that Type 1 New Entrants typically make considerable investments to enter the CC-Sector in order to help in the effort to meet the positive demand shock for CC-Products at the start of the pandemic. In order to recover these costs, new entrant firms ought to conduct a freedom-to-operate analysis as early as possible, particularly when following entry strategies (A) and (B), in order to mitigate the risk of infringement claims by incumbent firms pursuing a defensive counter strategy. Furthermore, when entering into cross-licensing agreements with Incumbent CC-Product Manufacturers, particularly in entry strategies (B) and (C), it would be prudent for Type 1 New Entrants to insist on a grant-back clause because it would allow them to retain ownership of their endogenously and newly developed CC-IP that potentially underlie valuable CC-Product enhancements that are desirable to incumbent firms. This, in turn, leaves new entrant firms enough leverage over incumbent firms to potentially negotiate lower royalties and less confining scope restriction clauses during the next round of cross-licensing negotiations.

This study comprises several contributions. We contribute theoretically to the IP management literature by offering a definition for a new IP-type that is relevant for the open innovation context, namely paraground IP. It is herein defined as IP that underlies innovations which are independently developed by an actor outside of a collaboration context on the basis of another (3rd party) actor's IP without authorised access or having established freedom-to-operate. Furthermore, we add methodologically by further enhancing and broadening the SVEL's scope of application to capturing IP dynamics in evolving innovation ecosystems. Finally, this study comprises managerial implications for decision makers and negotiators who represent incumbent and new entrant firms in CC-sectors as the crisis situation evolves.

The nature of the research methodology and the scarcity of publicly available data, however, add some limitations to our approach and resulting findings. As the future evolution of the Covid-19 pandemic and the resulting crisis-situation were unknown at the time this study was conducted, most of the managerial implications are based on consensus-reaching extrapolations from current events by subject matter experts. Nevertheless, we believe that our conclusions merit being empirically confirmed once the Covid-19 pandemic terminates and empirical data covering its entire evolution from onset to termination will be available. Finally, as the study is focused on the Covid-19 pandemic, the generalisability of our findings to other crisis situations remains to be tested. Future studies might want to apply the enhanced SVEL to other crisis settings, such as health or economic crises in order to either proof its versatility or to enhance and develop it further.

## References

4iP Council (2016) *Patents and inventive spirit at Dyson*. Available at: https://www.4ipcouncil.com/features/dyson-company-founded-inventor (Accessed: 10 October 2020).

Adner, R. and Kapoor, R. (2010) 'Value Creation in Innovation Ecosystems: How the Structure of Technological Interdependence Affects Firm Performance in New Technology Generations', *Strategic Management Journal*, 31, pp. 306–333.

Ambrosio, N. (2020) *Responding to the COVID-19 Crisis: Can Industry Help?*, *Four Twenty Seven*. Available at: http://427mgt.com/2020/03/26/responding-to-the-covid-19-crisis-can-industry-help/ (Accessed: 7 July 2020).

Auerswald, P. E. and Dani, L. (2017) 'The adaptive life cycle of entrepreneurial ecosystems: the biotechnology cluster', *Small Business Economics*. Small Business Economics, 49(1), pp. 97–117. doi: 10.1007/s11187-017-9869-3.

Azoulay, P. and Jones, B. (2020) 'Beat COVID-19 through innovation', *Science*, 368(6491), p. 553. doi: 10.1126/science.abc5792.

Basole, R. C. (2009) 'Visualization of interfirm relations in a converging mobile ecosystem', *Journal of Information Technology*, 24(2), pp. 144–159. doi: 10.1057/jit.2008.34.

Basole, R. C. *et al.* (2018) 'Ecoxight: Discovery, exploration, and analysis of business ecosystems using interactive visualization', *ACM Transactions on Management Information Systems*, 9(2). doi: 10.1145/3185047.

Bonakdar, A. *et al.* (2017) 'Capturing value from business models: the role of formal and informal protection strategies', *International Journal of Technology Management*, 73(4), p. 151. doi: 10.1504/IJTM.2017.10004002.

Bottomley, V. (2020) *Dyson's coronavirus response – Sir James Dyson on COVID-19 and developing the CoVent ventilator in 30 days*, *Odgers Berndtson Insights*. Available at: https://www.odgersberndtson.com/en-ch/insights/dysons-coronavirus-response-sir-james-dyson-on-covid-19-and-developing-the-covent-ventilator-in-30-days (Accessed: 14 August 2020).

Boudette, N. and Jacobs, A. (2020) *Inside G.M.'s Race to Build Ventilators, Before Trump's Attack*, *The New York Times*. Available at: https://www.nytimes.com/2020/03/30/business/gm-ventilators-coronavirus-trump.html (Accessed: 3 September 2020).

Brooks, C. and Flores, D. (2020) *Ventec Life Systems and GM Partner to Mass Produce Critical Care Ventilators in Response to COVID-19 Pandemic*, *GM Corporate Newsroom*. Available at: https://media.gm.com/media/us/en/gm/home.detail.html/content/Pages/news/us/en/2020/mar/0327-coronavirus-update-6-kokomo.html (Accessed: 3 September 2020).

Bruno, M. (2020) *Supply Chain Due For Overhaul After COVID-19, One Way Or Another*, *Aviationweek.com*. Available at: https://www.businesswire.com/news/home/20200724005112/en/Impact-of-COVID-19-on-the-Worlds-Automotive-Industry-2020---Strategies-Key-Market-Participants-are-Currently-

Employing---ResearchAndMarkets.com (Accessed: 11 June 2020).

Bryman, A. and Bell, E. (2015) *Business Research Methods*. 4th edn. Oxford, UK: Oxford University Press.

BusinessWire (2020) *Impact of COVID-19 on the World's Automotive Industry*, *ResearchAndMarkets.com*. Available at: https://www.businesswire.com/news/home/20200724005112/en/Impact-of-COVID-19-on-the-Worlds-Automotive-Industry-2020---Strategies-Key-Market-Participants-are-Currently-Employing---ResearchAndMarkets.com (Accessed: 11 October 2020).

Chatburn, R. L. (1991) 'A New System for Understanding Mechanical Ventilators', *Respiratory Care1*, 36(10), pp. 1123–1155.

Chen, M.-J. and Miller, D. (2015) 'Reconceptualizing competitive dynamics: A multidimensional framework', *Strategic Management Journal*, 36(5), pp. 758–775. doi: 10.1002/smj.2245.

Chesbrough, H. (2012) 'Open innovation: Where we've been and where we're going', *Research Technology Management*, 55(4), pp. 20–27. doi: 10.5437/08956308X5504085.

Contreras, J. L. *et al.* (2020) 'Pledging Intellectual Property for Covid-19', *Nature Biotechnology*, 38(October), pp. 1146–1150. doi: 10.1038/s41587-020-0682-1.

Elsen, M. *et al.* (2020) *Unpacking the innovation landscape for crisis-critical products, services and technologies in the Covid-19 pandemic*, *Innovation and Intellectual Property Management (IIPM) Lab*. Available at: https://www.iipm.eng.cam.ac.uk/news/unpacking-innovation-landscape-crisis-critical-products-services-and-technologies-covid-19 (Accessed: 13 October 2020).

F. Hoffmann-La Roche AG (2020) *Roche's cobas SARS-CoV-2 Test to detect novel coronavirus receives FDA Emergency Use Authorization*, *analytica-world.com*. Available at: https://www.analytica-world.com/en/news/1165398/roches-cobas-sars-cov-2-test-to-detect-novel-coronavirus-receives-fda-emergency-use-authorization.html (Accessed: 13 August 2020).

Ferguson, N. M. *et al.* (2020) *Impact of non-pharmaceutical interventions (NPIs) to reduce COVID-19 mortality and healthcare demand*, *Imperial College London*. doi: 10.25561/77482.

Garcia, R. and Calantone, R. (2002) 'A critical look at technological innovation typology and innovativeness terminology: a literature review', *The Journal of Product Innovation Management*, 19, pp. 110–132.

Girod, S. (2020) *How Luxury Brands Can Emerge Stronger From The Coronavirus Crisis*, *Forbes.com*. Available at: https://www.forbes.com/sites/stephanegirod/2020/07/31/how-luxury-brands-can-emerge-stronger-from-the-coronavirus-crisis/ (Accessed: 11 October 2020).

Granstrand, O. and Holgersson, M. (2013) 'Managing the intellectual property disassembly problem', *California Management Review*, 55(4), pp. 184–210. doi: 10.1525/cmr.2013.55.4.184.

Granstrand, O. and Holgersson, M. (2014) 'The challenge of closing open innovation: The intellectual property disassembly problem', *Research Technology Management*, 57(5), pp. 19–25. doi: 10.5437/08956308X5705258.

Granstrand, O. and Holgersson, M. (2020) ‘Innovation ecosystems: A conceptual review and a new definition’, *Technovation*, 90–91(May), p. 102098. doi: 10.1016/j.technovation.2019.102098.

Granstrand, O., Holgersson, M. and Opedal, A. (2020) ‘Towards fair pricing in technology trade and licensing’, *Stockholm Intellectual Property Law Review*, 3(1), pp. 6–15.

Hall, B. *et al.* (2014) ‘The Choice between Formal and Informal Intellectual Property: A Review’, *Journal of Economic Literature*, 52(2), pp. 375–423. doi: 10.1257/jel.52.2.375.

Herstatt, C. and Tiwari, R. (2020) ‘Opportunities of frugality in the post-corona era’, *International Journal of Technology Management*, 83(Nos. 1/2/3), pp. 15–33.

Hobday, M., Davies, A. and Prencipe, A. (2005) ‘Systems integration: A core capability of the modern corporation’, *Industrial and Corporate Change*, 14(6), pp. 1109–1143. doi: 10.1093/icc/dth080.

Holgersson, M. and Granstrand, O. (2017) ‘Patenting motives, technology strategies, and open innovation’, *Management Decision*, 55(6), pp. 1265–1284. doi: 10.1108/MD-04-2016-0233.

Holling, C. S. (1992) ‘Cross-Scale Morphology, Geometry, and Dynamics of Ecosystems’, *Ecological Monographs*, 62(4), pp. 447–502.

Jack, S. (2020) *Coronavirus: Government orders 10,000 ventilators from Dyson*, *BBC News*. Available at: https://www.bbc.com/news/business-52043767 (Accessed: 13 August 2020).

Justia (2020) *Patents by Inventor James Dyson*. Available at: https://patents.justia.com/inventor/james-dyson (Accessed: 10 October 2020).

Larrañeta, E., Dominguez-Robles, J. and Lamprou, D. A. (2020) ‘Additive Manufacturing Can Assist in the Fight against COVID-19 and Other Pandemics and Impact on the Global Supply Chain’, *3D Printing and Additive Manufacturing*, 7(3), pp. 100–103. doi: 10.1089/3dp.2020.0106.

López-Gómez, C. *et al.* (2020) *COVID-19 Critical Supplies: The Manufacturing Repurposing Challenge*, *United Nations Industrial Development Organization*. Available at: https://www.unido.org/news/covid-19-critical-supplies-manufacturing-repurposing-challenge (Accessed: 7 July 2020).

Ma, A. (2020) *NC textile mill ‘heeds call of nation,’ gears up to make 10 million face masks per week*, *The Charlotte Observer*. Available at: https://www.charlotteobserver.com/news/coronavirus/article241413386.html (Accessed: 13 August 2020).

Manero, A. *et al.* (2020) ‘Leveraging 3D printing capacity in times of crisis: Recommendations for COVID-19 distributed manufacturing for medical equipment rapid response’, *International Journal of Environmental Research and Public Health*, 17(13), pp. 1–17. doi: 10.3390/ijerph17134634.

Moerchel, A., Tietze, F. and Urmetzer, F. (2021) *[Forthcoming]. Mapping Competitive Dynamics in Business Ecosystems - A Case Study from the Aerospace Industry*. Cambridge, UK.

Moore, J. F. (1993) ‘Predators and prey: a new ecology of competition.’, *Harvard Business*

*Review*, 71(3), pp. 75–86.

Nicholson Price II, W., Rai, A. K. and Minssen, T. (2020) 'Knowledge transfer for large-scale vaccine manufacturing', *Science*, 369(6506), pp. 912–914.

Nicola, M. *et al.* (2020) 'The Socio-Economic Implications of the Coronavirus and COVID-19 Pandemic: A Review', *International Journal of Surgery*. Elsevier, 78(March), pp. 185–193. doi: 10.1016/j.ijsu.2020.04.018.

Oxford Business Group (2020) *The impact of Covid-19 on global supply chains*, *Covid-19 Economic Assessments*. Available at: https://oxfordbusinessgroup.com/news/impact-covid-19-global-supply-chains (Accessed: 11 June 2020).

Pays, C. (2020) *The beauty industry is rallying to fight Covid-19*, *Vogue*. Available at: https://www.vogue.fr/beauty-tips/article/lvmh-loreal-beauty-companies-producing-hand-sanitizer-to-help-hospitals (Accessed: 13 August 2020).

Phillips, M. A. and Srai, J. S. (2018) 'Exploring emerging ecosystem boundaries: Defining "the game"', *International Journal of Innovation Management*, 22(8), pp. 1–21. doi: 10.1142/S1363919618400121.

Phua, J. *et al.* (2020) 'Intensive care management of coronavirus disease 2019 (COVID-19): challenges and recommendations', *The Lancet Respiratory Medicine*. Elsevier Ltd, 8(5), pp. 506–517. doi: 10.1016/S2213-2600(20)30161-2.

Poltorak, A. and Lerner, P. (2011) *Essential of Intellectual Property*. 2nd Editio, *Development*. 2nd Editio. Hoboken, New Jersey: John Wiley & Sons.

Richards, G. (2020) *Project Pitlane: how rival F1 teams united in battle against Covid-19*, *The Guardian*. Available at: https://www.theguardian.com/sport/2020/may/24/project-pitlane-how-rival-f1-teams-united-in-battle-against-covid-19 (Accessed: 13 August 2020).

Robson, C. and McCartan, K. (2017) *Real World Research*. 4th edn. Chichester, UK: John Wiley & Sons Ltd.

Samuelson, P. (1941) 'The Stability of Equilibrium: Comparative Statics and Dynamics', *Econometrica*, 9(2), pp. 97–120.

Sohrabi, C. *et al.* (2020) 'World Health Organization declares global emergency: A review of the 2019 novel coronavirus (COVID-19)', *International Journal of Surgery*. Elsevier, 76(February), pp. 71–76. doi: 10.1016/j.ijsu.2020.02.034.

The James Dyson Foundation (2020) *Our Story: The accidental engineer*, *The James Dyson Foundation,*. Available at: https://www.jamesdysonfoundation.com/who-we-are/our-story.html (Accessed: 13 August 2020).

Tietze, F. *et al.* (2020) 'Crisis-Critical Intellectual Property: Findings From the COVID-19 Pandemic', *IEEE Transactions on Engineering Management*, pp. 1–18. doi: 10.1109/TEM.2020.2996982.

Tiwari, R. and Herstatt, C. (2012) 'Assessing India's lead market potential for cost-effective innovations', *Journal of Indian Business Research*, 4(2), pp. 97–115. doi: 10.1108/17554191211228029.

Urmetzer, F., Gill, A. and Reed, N. (2018) 'Using Business Ecosystem Mapping to Generate New Competitive Value Propositions', in *CIE48*. Auckland: Computers and Industrial

Engineering.

Van De Ven, A. H. and Delbecq, A. L. (1974) 'The Effectiveness of Nominal, Delphi, and Interacting Group Decision Making Processes', *Academy of Management Journal*, 17(4), pp. 605–621. doi: 10.5465/255641.

Weyrauch, T. and Herstatt, C. (2017) 'What is frugal innovation? Three defining criteria', *Journal of Frugal Innovation*. Journal of Frugal Innovation, 2(1), pp. 1–17. doi: 10.1186/s40669-016-0005-y.

World Health Organisation (2003) *SARS outbreak contained worldwide*, *Media centre*. Available at: http://www.who.int/mediacentre/news/releases/2003/pr56/en/ (Accessed: 16 June 2020).

World Health Organisation (2020a) *Rational use of personal protective equipment for coronavirus disease 2019 (COVID-19) and considerations during severe shortages*. Available at: https://www.who.int/publications/i/item/rational-use-of-personal-protective-equipment-for-coronavirus-disease-(covid-19)-and-considerations-during-severe-shortages.

World Health Organisation (2020b) *Situation Report*, *Coronavirus disease COVID-2019*. Available at: https://www.who.int/emergencies/diseases/novel-coronavirus-2019/situation-reports/.

World Health Organisation (2020c) *WHO Director-General's opening remarks at the media briefing on COVID-19*, *World Health Organization*. Available at: https://www.who.int/dg/speeches/detail/who-director-general-s-opening-remarks-at-the-media-briefing-on-covid-19---11-march-2020 (Accessed: 28 July 2020).

Woudenberg, F. (1991) 'An evaluation of Delphi', *Technological Forecasting and Social Change*, 40(2), pp. 131–150. doi: 10.1016/0040-1625(91)90002-W.

Yasiejko, B. C. (2020) *Regeneron Antibody Cocktail Used by Trump Faces Patent Suit*, *Bloomberg Law*. Available at: https://news.bloomberglaw.com/pharma-and-life-sciences/regeneron-antibody-cocktail-used-by-trump-faces-patent-suit (Accessed: 9 October 2020).